\documentclass[%
longbibliography,%
 sor4-1,%
 amsmath,amssymb,
 reprint,%
]{revtex4-1}

\usepackage{natbib}

\usepackage{graphicx}
\usepackage{dcolumn}
\usepackage{bm}
\usepackage{color,amsmath,amssymb}
\usepackage[normalem]{ulem}

\newcommand{\gdot}{\ensuremath{\dot{\gamma}}}
\newcommand{\be}{\begin{equation}}
\newcommand{\ee}{\end{equation}}
\newcommand{\beqna}{\begin{eqnarray}}
\newcommand{\eeqna}{\end{eqnarray}}

\newcommand{\cycband}{\Delta_c\gdot}
\newcommand{\maxband}{\Delta_m\gdot}
\newcommand{\at}[2][]{#1|_{#2}}
\newcommand{\tw}{t_{\rm w}}
\newcommand{\Gc}{G_{\rm c}}
\newcommand{\sigmay}{\sigma_{\rm y}}
\newcommand{\sigmadyn}{\sigma_{\rm dyn}}
\newcommand{\sigmamax}{\sigma_{\rm max}}
\newcommand{\gammaac}{\gamma_{\rm ac}}

\newcommand{\tauinv}{\ensuremath{\frac{1}{\tau}}}
\newcommand{\crcor}{\ensuremath{\langle\hat{\rho}_{\gdot,\tauinv}(y)\rangle}}

\begin{document}
\preprint{AIP/123-QED}

\title{Shear banding in large amplitude oscillatory shear (LAOStrain and LAOStress) of\newline  soft glassy materials}

\author{Rangarajan Radhakrishnan}%
\author{Suzanne M. Fielding}%
\affiliation{%
  Department of Physics, Durham University, Science Laboratories,
  South Road, Durham DH1 3LE, UK} \date{\today}

\date{\today}
\begin{abstract}
  We study theoretically shear banding in soft glassy materials subject to large amplitude time-periodic shear flows, considering separately the protocols of large amplitude oscillatory shear strain, large amplitude square or triangular or sawtooth strain rate, and large amplitude oscillatory shear stress. In each case, we find shear banding to be an important part of the material's flow response for a broad range of values of the frequency $\omega$ and amplitude of the imposed oscillation. Crucially, and highly counterintuitively, in the glass phase this persists even to the lowest frequencies accessible numerically (in a manner that furthermore seems consistent with it persisting even to the limit of zero frequency $\omega\to 0$), even though the soft glassy rheology model in which we perform our calculations has a purely monotonic underlying constitutive curve of shear stress as a function of shear rate, and is therefore unable to support shear banding as its true steady state response to a steadily imposed shear of constant rate. We attribute this to the repeated competition, within each flow cycle, of glassy aging and flow rejuvenation. Besides reporting significant banding in the glass phase, where the flow curve has a yield stress, we also observe it at noise temperatures just above the glass point, where the model has a flow curve of power law fluid form.  In this way, our results suggest a predisposition to shear banding in flows of even extremely slow time-variation, for both aging yield stress fluids, and for power law fluids with sluggish relaxation timescales. We show that shear banding can have a pronounced effect on the shape of the Lissajous-Bowditch curves that are commonly used to fingerprint complex fluids rheologically.  We therefore counsel caution in seeking to compute such curves in any calculation that imposes upfront a homogeneous shear flow, discarding the possibility of banding.  We also analyze the stress response to the imposed strain waveforms in terms of a `sequence of physical processes'.
\end{abstract}

\pacs{Valid PACS appear here}
\maketitle

\section{Introduction}
\label{sec:introduction}

A broad class of disordered soft materials, including emulsions~\cite{Becu:2006}, foams~\cite{Rouyer2008}, colloids~\cite{Mason1995,Knaebel2000}, microgels~\cite{Cloitre2000}, and star polymers~\cite{Rogers2010}, share in common several notable rheological properties.  In nonlinear flows, their steady state flow curve of shear stress $\sigma$ as a function of shear rate $\gdot$ is often fit to the form $\sigma=\sigmay+a\gdot^n$ with $n<1$, corresponding to yield stress fluid behavior for $\sigmay> 0$ and power law fluid behavior for $\sigmay=0$. In the regime of linear response, under a small amplitude oscillatory shear strain, their viscoelastic storage and loss moduli, $G'(\omega)$ and $G''(\omega)$, are often in near constant ratio, with $G''/G'$ typically about $0.1$, and with both functions showing only a weak or negligible frequency dependence down to the lowest accessible frequencies.

Consistent with the existence of these sluggish relaxation modes, another striking feature is that of rheological aging~\cite{Fielding:2000}, in which a sample's flow response becomes progressively more solid-like as a function of its own age $\tw$, defined as the time since it was freshly prepared at time $t=0$, for example by loading it into a rheometer and preshearing it, before a test deformation is later applied after a waiting time $t=\tw$.  The application of a sustained shear flow will however typically halt this aging process and rejuvenate the sample to a steady state with an effective age set by the inverse flow rate $1/\gdot$.

These shared rheological features have been attributed to the generic presence in these materials of the underlying `glassy' features of structural disorder ({\it e.g.} in a disordered packing of emulsion droplets or foam bubbles) and metastability ({\it e.g.}  in the large energy barriers involved in stretching soap films, which impede droplet rearrangements). The term `soft glassy materials' has accordingly been coined to describe them~\cite{sollich1997,Fielding2014}.

In the rheological literature, soft glasses are often also referred to as yield stress fluids. Recently, these have been suggested to fall into two broad categories: `simple' and `viscosity bifurcating'~\cite{Moller2009,Fielding2014} yield stress fluids. Among these, viscosity bifurcating fluids~\cite{Ragouilliaux2007,Moller2009,Fall2010,Martin2012} typically exhibit a strong time dependence (sometimes called thixotropy) in their transient rheological response. Furthermore, under a sustained applied shear flow they typically exhibit shear banding, with their steady state flow field comprising macroscopic bands of differing viscosities, with layer normals in the flow-gradient direction. This ability to support steady state shear bands is thought to stem from a non-monotonicity in the underlying constitutive curve of shear stress as a function of shear rate (as pertaining to initially homogeneous flow states).  In contrast, simple yield stress fluids~\cite{Coussot2009,Ovarlez2010,Ovarlez2013} typically show much weaker thixotropy and are thought to have a monotonic constitutive curve, being thereby incapable of exhibiting shear banding as their steady response to a sustained applied shear flow (at least in the absence of concentration coupling).

Beyond the steady state shear banding just described, recent years have seen an increasing realization that shear bands might also form quite generically in flows that involve a strong time-dependence~\cite{Fielding:2016,Moorcroft2011}, 
even in materials that have a purely monotonic underlying constitutive curve and are therefore incapable of supporting shear bands as their steady state response to a sustained applied shear flow of constant rate. (In fact this prediction applies not only to soft glassy materials but to complex fluids more generally~\cite{Moorcroft2013a,Fielding2014}, though we restrict our attention to soft glasses in this work.) To date, this concept has been investigated in detail in the transiently time-dependent flows of shear startup and step stress, as we now summarize.

In shear startup, an initially well rested sample is subject at some time $t=\tw$ to the switch-on of a shear rate $\gdot$ that is held constant thereafter. Measured in response to this is the material's shear stress startup curve as a function of the time (or equivalently of the accumulated strain) since the inception of the flow. Typically, this signal rises initially linearly at early times, before displaying an overshoot after which the stress finally falls to attain its steady state value as prescribed by the material's flow curve at the given imposed shear rate.

In Ref.~\cite{Moorcroft2013a}, it was suggested that the presence of this overshoot should generically predispose a material to the formation of shear bands, at least transiently, as the stress declines from its overshoot to the final steady state value.  (In this steady state, the flow field may either remain banded, in a viscosity bifurcating fluid; or heal back to homogeneous flow, in a simple yield stress fluid with a monotonic underlying constitutive curve.)  This phenomenon has indeed been widely observed: experimentally in carbopol gel~\cite{divoux2010,Divoux:2011b}, Laponite clay~\cite{Martinetal2012a,gibaud2008}, a non-Brownian fused silica suspension~\cite{Kurokawa:2015} and waxy crude oil~\cite{Dimitriou:2014}; in molecular simulations of a colloidal gel~\cite{Colombo:2014}, polymeric fluids~\cite{Mohaghehi:2016} and molecular glasses~\cite{Yunfeng:2007,Varnik:2004}; and in theoretical studies of a model foam~\cite{Kabla:2007,Barry:2010}, the soft glassy rheology and fluidity models~\cite{Moorcroft2011,Fielding2014,Lehtinen:2013}, the STZ model of amorphous elastoplastic solids~\cite{Manning:2007,Manning:2009a}, a mesoscopic model of plasticity~\cite{Jagla:2010}, and a model of polymer glasses~\cite{Fielding:2013}. In cases where the height of the stress overshoot increases as a function of the age of the sample before shearing commenced, the severity of the shear banding is predicted to increase accordingly.

In a step stress experiment, an initially well rested sample is subject at some time $t=\tw$ to the switch-on of a constant stress $\sigma$ that is held constant thereafter. Measured in response to this is the material's creep curve $\gamma(t)$, often reported as its time-differential $\gdot(t)$.  In soft glasses, this signal typically displays an initial regime of slow creep in which $\gdot$ progressively decreases over time, followed (for stress values $\sigma>\sigmay$) by a yielding process in which $\gdot$ increases to finally attain its value as prescribed by the steady state flow curve at the given stress. In Ref.~\cite{Moorcroft2013a}, it was suggested that a material should be generically predisposed to the formation of shear bands during this yielding process that follows the initial regime of slow creep, during the time-interval over which the time-differentiated creep curve simultaneously curves up and slopes up as a function of time and the sample starts flowing.  This phenomenon has indeed been observed: experimentally in carbopol gel~\cite{Divoux:2011a,Magnin:1990}, carbon black~\cite{Gibaud:2010,Grenard:2014} and a colloidal glass~\cite{Sentja:2015}; in particle based simulations of colloidal glasses~\cite{Chaudhuri:2013}; and in stochastic simulations of the soft glassy rheology model~\cite{Moorcroft2013a,Fielding2014}.

In the shear startup and step stress protocols just described, the time-dependence is transient in nature, typically persisting for just a few strain units during the time taken to establish a final steady flow out of an initial rest state.  In consequence, for a simple yield stress fluid at least, the associated shear banding is itself transient: the bands that form as the material initially yields and starts flowing then subsequently heal away to give a homogeneous final steady state.  (A viscosity bifurcating fluid can instead maintain bands even in steady state, due to the non-monotonic underlying constitutive curve.)

In view of this, an important question of fundamental principle is whether an imposed flow that has a {\em sustained} time-dependence can give rise to correspondingly sustained shear banding, even in a simple yield stress fluid that is unable to support banding as its ultimate steady state response to a steadily imposed shear flow of constant rate.  Indeed, intuitively we might expect a square wave caricature of a large amplitude oscillatory strain to correspond to a repeating sequence of forward then reverse shear startup runs. In any regime in which these repeated startup events are associated with an overshoot in the signal of stress as a function of strain, we might intuitively expect shear banding in each half cycle, associated with these overshoots. Likewise, we might intuitively expect a square wave caricature of a large amplitude oscillatory stress to correspond to a repeated sequence of positive then negative step stress experiments.  In any regime in which each repeated step is associated with a yielding process of the kind discussed above for the simpler protocol of a single step stress, we might intuitively expect to find shear banding associated with these yielding events in each half cycle.

In what follows, we investigate this scenario by studying the response of the soft glassy rheology (SGR) model~\cite{sollich1997,Sollich1998},
in its form as extended to allow for the possibility of heterogeneous shear flows~\cite{Fielding2008},
to several different large amplitude time-periodic imposed shear flows. We consider in turn the protocols of large amplitude oscillatory shear strain (LAOStrain), large amplitude square wave strain rate, large amplitude triangle wave strain rate, large amplitude sawtooth strain rate, and large amplitude oscillatory shear stress (LAOStress).  In each case, we shall demonstrate shear banding to be an important part of the flow response across a wide range of values of the amplitude $\gamma_0$ (or $\sigma_0$) and frequency $\omega$ of the imposed flow.

In the limit of zero frequency $\omega\to 0$ of the imposed oscillation, our initial intuition might lead us to expect to recover a situation in which the system simply quasistatically sweeps up and down its steady state flow curve during the course of each cycle, with the flow remaining homogeneous at all times (in a simple yield stress fluid at least).  Crucially, however - and counterintuitively - in the glass phase we shall find that banding persists even at the lowest frequencies accessible numerically,  in a manner that furthermore appears consistent with the idea that it would persist even to  the limit $\omega\to 0$, were this accessible numerically.  We emphasize that this is true even for the simple yield stress fluids considered here, which have a purely monotonic underlying constitutive curve and are unable to support banding as their true steady state response to a sustained applied shear of constant rate $\gdot$.  We shall show that this arises from a repeated competition, within each cycle, between glassy aging and flow-rejuvenation: the sample ages (with its typical stress relaxation timescale $\tau$ increasing) during the weak flow phase of each cycle, then is rejuvenated during the strong flow phase (with $\tau$ decreasing).  Put simply: an aging material has no fixed intrinsic stress relaxation rate $1/\tau$ compared to which we can set the driving frequency $\omega$ to be small and expect to recover steady state response. This scenario has far reaching implications for the flow behavior of aging glassy materials, in suggesting a possible generic predisposition to shear banding even in flows of arbitrarily slow time-dependence.

The protocol of large amplitude oscillatory shear (LAOS)~\cite{Hyun2011} has been the focus of intense interest in the rheology community in recent years, in particular for its suggested use in `fingerprinting' complex fluids via a series of tests in which the amplitude and frequency of the imposed oscillation are separately varied. At high frequencies, a material's elastic response is probed.  At low frequencies, viscous response might a priori be expected (although in the aging materials of interest here that idea should be treated with caution in view of the remarks of the previous paragraph).  Large amplitudes flows probe nonlinear response, while linear viscoelastic response is recovered for small amplitudes.

\begin{figure*}[!tbp]
\includegraphics[width=15.0cm]{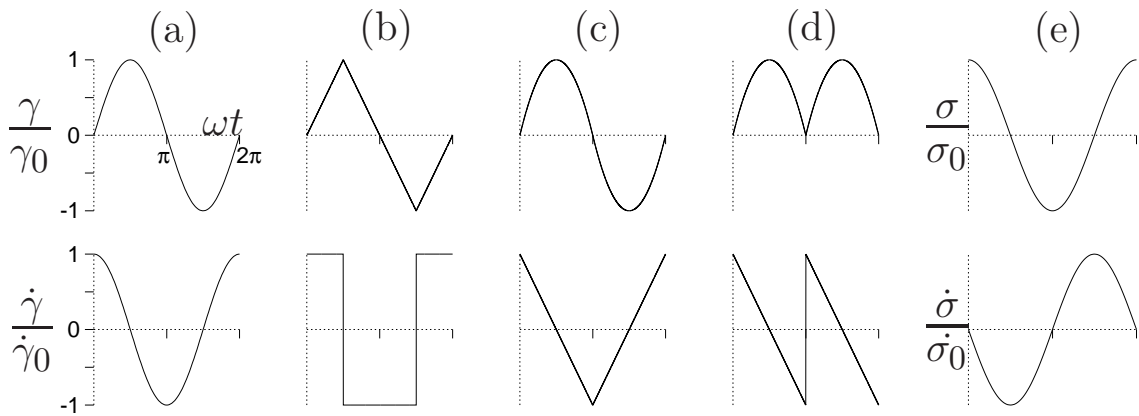}
\caption{The large amplitude time-periodic shear flows that we shall
  consider: a) oscillatory strain, b) square wave strain rate, c) triangle wave strain rate, d) sawtooth strain rate, and e) oscillatory stress. For each of a) to e), the top panel shows the strain (or stress) and the bottom panel shows the corresponding rate. The horizontal axis is the same in each subpanel.}
\label{fig:protocols}
\end{figure*}

In the context of yield stress fluids, LAOS has been studied both experimentally~\cite{Yoshimura1987,Knaebel2000,Viasnoff2003,Rouyer2008,Ewoldt2010,Renou2010,Guo2011,VanderVaart2013,Koumakis2013,Poulos2013,Poulos2015} and theoretically~\cite{Yoshimura1987,Viasnoff2003,Ewoldt2010,Rogers:2011,Rogers:2012,Koumakis2013,Mendes2013,Blackwell2014,Sollich1998,Rouyer2008}. In terms of a consideration of shear banding in this protocol, however, few experiments have directly imaged the flow field across the sample, although strain localization was reported in foam in Ref.~\cite{Rouyer2008} and in concentrated suspensions in Ref.~\cite{Guo2011}.  In similar spirit, all the theoretical studies of which we are aware have simply assumed the flow to remain homogeneous, discarding upfront the possibility of banding.  

A central contribution of this work is to suggest that aging yield stress fluids might generically be expected to exhibit shear banding in LAOS, and furthermore that the presence of banding has a major influence on the measured bulk rheological signals.  Indeed, we shall show that a system's Lissajous-Bowditch curves can differ strongly when calculated within the assumption of a purely homogeneous flow, compared with a calculation that allows bands to form.  This suggests that attempts to rheologically fingerprint a fluid without taking banding properly into account -- as is widespread in much of the existing theoretical LAOS literature -- should be treated with caution.

In a previous Letter~\cite{Ranga:2016}, we announced the basic result that an aging yield stress fluid, as modeled by the soft glassy rheology model in its glass phase, can exhibit shear banding in large amplitude time-periodic shear strain protocols. That study was restricted to the model's glass phase, where its noise temperature parameter (defined below) $x<1$, presenting numerical results for the single value $x=0.3$.  The present paper contains a much more detailed discussion of the results announced in Ref.~\cite{Ranga:2016}. It also extends our study to a much broader range of noise temperatures, including those above the glass point, $x>1$, where the model shows power law fluid behavior, with no yield stress.  We report significant banding here too, suggesting that the scenario is applicable not only to aging yield stress fluids, but also to power law fluids with sluggish relaxation timescales.  The present manuscript also gives new results for shear banding of soft glasses in large amplitude oscillatory stress.  

The paper is structured as follows. In Sec.~\ref{sec:protocols} we define the flow protocols to be considered.  Sec.~\ref{sec:model} outlines the SGR model in which we shall perform the study, together with our simulation method and some results used to benchmark it. We then present our results: in Sec.~\ref{sec:LAOStrain} for shear banding in large amplitude oscillatory shear strain, in Sec.~\ref{sec:LAOSothers} for large amplitude square or triangular or sawtooth wave strain rates, and in Sec.~\ref{sec:LAOStress} for large amplitude oscillatory shear stress. Sec.~\ref{sec:conclusions} discusses our conclusions.

\section{Flow Protocols}
\label{sec:protocols}

In this section, we define the rheological protocols to be studied throughout the paper. In each case, we shall consider a sample of fluid that is freshly prepared at some time $t=0$ and then left to age undisturbed for a waiting time $\tw$ before the periodic flow is switched on. (We shall discuss in Sec.~\ref{sec:model} the way in which we model a freshly prepared sample in the SGR model.) 

For the imposed flow, we shall consider several different possible waveforms, listed as follows. For each strain-imposed waveform, the strain amplitude will be denoted $\gamma_0$, and the strain-rate amplitude $\gdot_0$. Likewise in the stress-imposed waveform, the stress amplitude is denoted $\sigma_0$ and the  amplitude of the rate of change of the stress $\dot{\sigma}_0$.
\begin{itemize}

\item Large amplitude oscillatory shear strain, abbreviated to LAOStrain. Here $\gamma(t)=\gamma_0\sin(\omega (t-\tw))$. See Fig.~\ref{fig:protocols}a).

\item Large amplitude square wave strain rate, in which the strain rate periodically switches between equal positive and negative values, with a switching time $\pi/\omega$. The associated strain signal is triangular. See Fig.~\ref{fig:protocols}b).

\item Large amplitude triangular wave strain rate, in which the strain rate is piecewise linear and continuous in value but with repeated slope discontinuities. See Fig.~\ref{fig:protocols}c).

\item Large amplitude sawtooth wave strain rate, in which the strain rate is piecewise linear with repeated discontinuities in value.  See Fig.~\ref{fig:protocols}d).

\item Large amplitude oscillatory shear stress, abbreviated to LAOStress. Here $\sigma=\sigma_0 \cos(\omega (t-\tw))$. See Fig.~\ref{fig:protocols}e).

\end{itemize}

After many cycles have been performed, in any regime where significant shear banding arises, the response of the system becomes (at least to excellent approximation) invariant from cycle-to-cycle $t\to t+2\pi/\omega$, and independent of the waiting time $\tw$ before the flow commenced.  For an initial waiting time $\tw=10.0$, this state of cycle-to-cycle invariance is typically achieved after $50$ cycles.

Except where stated, all our results below are for an initial waiting time $\tw=10.0$ and for a run in which $50$ cycles are performed before we then start taking measurements. Such results have therefore achieved cycle-to-cycle invariance. Indeed, to obtain better statistics in calculating the Lissajous-Bowditch curves, we generally average the data over the $50$th to $100$th cycles.

An entirely feasible experimental protocol, however, would be to wait for the sample to become highly aged before then performing a LAOS run comprising just a few tens of cycles.  Accordingly, we shall also present data for $\tw\to\infty$ (i.e., initializing the sample in equilibrium above the glass point), performing only $50$ cycles before we then average the system's response over the next $50$ cycles. (During those $50$ cycles over which we average, a small degree of time-variation does in fact occur, during the system's slow transient evolution to the state of cycle-to-cycle invariance after $1000$ cycles.)  Such data are clearly not in a state of cycle-to-cycle invariance, but do correspond to the experimentally feasible situation of an old sample subject to a few tens of LAOS cycles.

At lower strain amplitudes, in the absence of shear banding, indefinite cycle-to-cycle aging is expected even after many cycles.  This has been studied in detail previously~\cite{Fielding:2000} and we do not consider it further here.

To seed the formation of shear bands we add a small perturbation to the initial condition, such that the effective initial sample age as a function of position $y$ across the rheometer gap of width $L_y$ is $\tw\left[1+\epsilon\cos(2\pi y/L_y)\right]$ with $\epsilon=0.1$. In obtaining the result of Fig~\ref{fig:transition} only, we also (in order to mitigate noise) included a toy model of flow cell curvature, by rendering the shear stress  a function of position across the cell $\sigma\left[1+\kappa\cos(2\pi y/L_y)\right]$ with $\kappa=0.01$.  (In true planar shear, the stress stress must be uniform across the cell, giving $\kappa=0$.)

\section{Soft Glassy Rheology Model}
\label{sec:model}
We perform our study within the soft glassy rheology (SGR) model, which we now summarize, referring the reader to Refs.~\cite{sollich1997,Sollich1998,Fielding2008} for full details. The model considers an ensemble of elements, each of which is taken to correspond to a local mesoscopic region of a soft glassy material comprising (say) a few tens of emulsion droplets.  Each element is assigned local continuum variables of shear strain $l$ and stress $kl$, with $k$ constant, which describe the elastic deformation of this region of material relative to a state of locally undeformed equilibrium.  The macroscopic stress of the sample as a whole is taken to be the average over the local elemental stresses:
\be
\sigma(t)=k\int dE \int dl\; l P(E,l,t).
\ee

The elements are then taken to undergo loading and activated hopping dynamics in an energy landscape of traps, as follows. Under an imposed deformation, each element experiences a buildup of local elastic stress such that, between hops, the local intra-trap strain of each element affinely follows the macroscopic strain field, $\dot{l}=\gdot$. These local stresses are then intermittently released by local plastic yielding events. Each such yielding event is taken to correspond to the hopping of an element out of one trap and into another. These hopping events are modeled as being dynamically activated: an element in a trap of depth $E$ and with local shear strain $l$ is assigned a probability per unit time of yielding given by $\tau^{-1}(E,l)=\tau_0^{-1}\exp\left[-(E-\tfrac{1}{2}kl^2)/x\right]$.  In this expression, the parameter $x$ is an effective mean field noise temperature that is intended to model in a mean field way coupling with other yielding events elsewhere in the sample.  Upon yielding, an element instantaneously resets its local stress to zero and selects its new energy barrier at random from a distribution $\rho(E)=\exp(-E/x_g)$.  In a freshly prepared sample, we assume a distribution $P(E,l)=\rho(E)\delta(l)$, corresponding to a well rested system just quenched from a high noise temperature.

This exponential `prior' distribution $\rho(E)$ confers a broad spectrum of yielding times $P(\tau)$ and results in a glass phase for $x<x_g$ in which the model exhibits rheological aging, with the typical relaxation timescale increasing linearly with the system's age $\tw$ in the absence of flow.  The application of a sustained flow however rejuvenates the sample and restores it to an effective age that is set by the inverse flow rate $1/\gdot$.  Throughout we use units in which $x_g=1$, $k=1$ and $\tau_0=1$.

The steady state flow curve $\sigma(\gdot)$ of shear stress as a function of shear rate has a yield stress $\sigmay(x)$ for noise temperature $x<1$ in the glass phase, beyond which it rises monotonically according to $\sigma-\sigmay\sim \gdot^{1-x}$. This gives simple yield stress fluid behavior, precluding steady state banding.  For noise temperatures $1<x<2$ the flow curve is of power law fluid form, with $\sigma\sim\gdot^{x-1}$. For $x>2$, we recover a Newtonian flow curve with $\sigma\sim \gdot$.

\begin{figure}[!tbp]
  \includegraphics{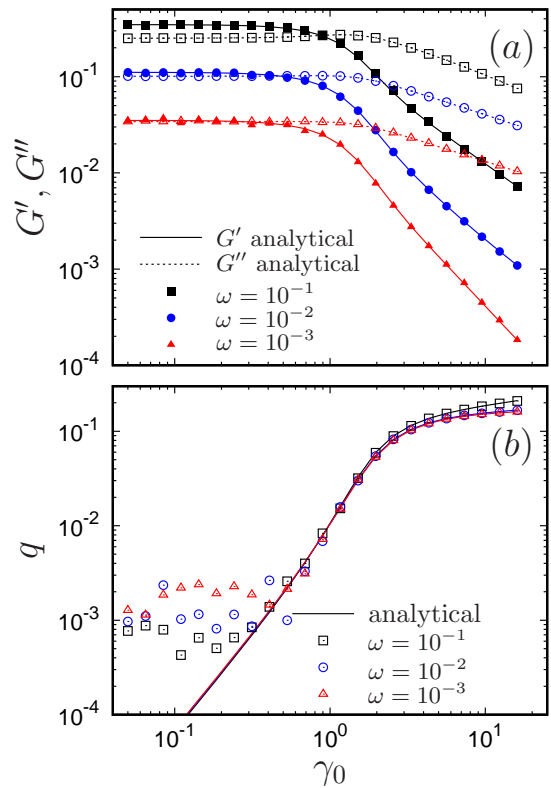} \caption{Results of our
  waiting-time Monte Carlo simulations (symbols) for the homogeneous
  form of the SGR model subject to large amplitude oscillatory shear
  strain, compared with independent results for the same quantities
  obtained from analytical expressions (lines). Panel (a) shows the
  storage $G'$ (filled symbols) and loss $G''$ (unfilled symbols)
  modulus for the fundamental mode; and (b) shows the residual $q$
  measuring the weight in all higher modes.  For each quantity, curves
  top to bottom are for frequency values $\omega=10^{-1}
  (\square),10^{-2} (\textcolor{blue}{\circ}),10^{-3}
  (\textcolor{red}{\triangle})$.  The noise temperature $x=1.5$, above
  the glass transition. Number of streamlines $n=1$, number of SGR
  elements per streamline $m=1000$. We thank Prof.  Peter Sollich for
  providing us with the data from the analytical
  expressions~\cite{Sollich1998}.}
\label{fig:comp_sollich}
\end{figure}
So far we have described the model in its original form~\cite{sollich1997,Sollich1998}, which is spatially homogeneous and unable to account for any heterogeneous flow effects such as shear banding. In Refs.~\cite{Fielding2008,Moorcroft2013a}, we provided an extension to the model to allow for the formation of shear bands coexisting with layer normals in the flow gradient direction $y$. This adopts a 1D approach in which the velocity is confined to the flow direction $x$ and varies only in the flow-gradient direction $y$, with the $y$ coordinate discretized into $i=1 \cdots n$ streamlines of equal spacing $L_y/n$, for a sample of thickness $L_y$ between rheometer plates at $y=0,L_y$.  The shear rate field is then $\gdot_i(t)$ as the coordinate $y$ varies across the streamlines $i=1\cdots n$. At any streamline $i$, this is related to the fluid velocity $v$ in the $x$ direction by the spatially discretized derivative $\gdot(y)=dv(y)/dy$, i.e., $\gdot_i=(v_{i+1}-v_{i-1})/(2L_y/n)$.

Although the shear rate field does not vary in $x$, each streamline has its own sub-ensemble of $j=1\cdots m$ SGR elements, with the shear stress of the $i$th streamline defined as $\sigma_i=(k/m)\sum_j l_{ij}$. In this way, this 1D model essentially comprises a series of SGR models stacked in the $y$ direction, coupled by a 1D Stokesian force balance, which we now describe.

In zero Reynolds number conditions of creeping flow, which we assume throughout, the condition of force balance imposes, in this 1D approach, that the shear stress must remain uniform across all streamlines at all times, $\sigma_i(t)=\sigma(t)$.  However, suppose a hop occurs at element $ij$ when its local strain is $l=\ell$, reducing the stress on that streamline. With the model as described so far, this potentially violates force balance. To correct for this, we restore force balance by updating all elements on the same streamline $i$ according to $l\to l + \ell/m$. This ensures uniform stress across streamlines, but with an overall sample stress that is incorrectly unchanged compared with that before the hop.  To ensure a properly reduced global stress after the hop, we then update all elements on all streamlines as $l \to l - \ell/mn$.

The scenario of force balance just described implements the propagator implied by Stokesian balance in the single spatial dimension $y$, with translational invariance imposed in $x$. A 2D approach would instead be possible, using the 2D propagator discussed in detail in~\cite{Picard:2004} and used in 2D elasto-plastic lattice models in~\cite{Picard:2005}. (Indeed Ref.~\cite{Picard:2004} describes how its 2D propagator reduces to 1D upon integrating over the flow direction $x$.) We expect our 1D approach to be well suited to the problem in hand here, of studying shear bands that form with layer normals in the flow gradient direction.

To account for the structure of the interface between any shear bands that form~\cite{Lu1999}, we further incorporate a small stress diffusivity between neighboring streamlines. To do so, after the hop of an element with strain $\ell$ on streamline $i$ as just described, we further adjust the strain of three randomly chosen elements on each adjacent streamline $i\pm 1$ by $\ell w(-1,+2,-1)$, with $w$ small.

Our numerical simulations of this model are performed using an event-driven waiting time Monte Carlo algorithm~\cite{Bortz:1975,Gillespie:1976,Fielding2008}.  In each `event', the next element to yield is selected stochastically: the probability $P_{ij}$ that the next element to yield is the $j$th particle on the $i$th streamline is $P_{ij}=r_{ij}/\sum_{ij}r_{ij}$, given an elemental hop rate $r_{ij}=\tau^{-1}(E_{ij},l_{ij})=\tau_0^{-1}\exp\left[-(E_{ij}-\tfrac{1}{2}kl_{ij}^2)/x\right]$.  The time interval $dt$ to the next hop is also selected in a
stochastic way: $dt=-\ln(s)/\sum_{ij}r_{ij}$, where $s$ is a random number selected from a uniform distribution between $0$ and $1$. All results reported are converged with respect to increasing the number of streamlines $n$ and the number of elements per streamline $m$. For further details of this simulation method, the reader is referred to~\cite{Fielding2008}.

\begin{figure*}[!tbp]
\includegraphics[width=18.0cm]{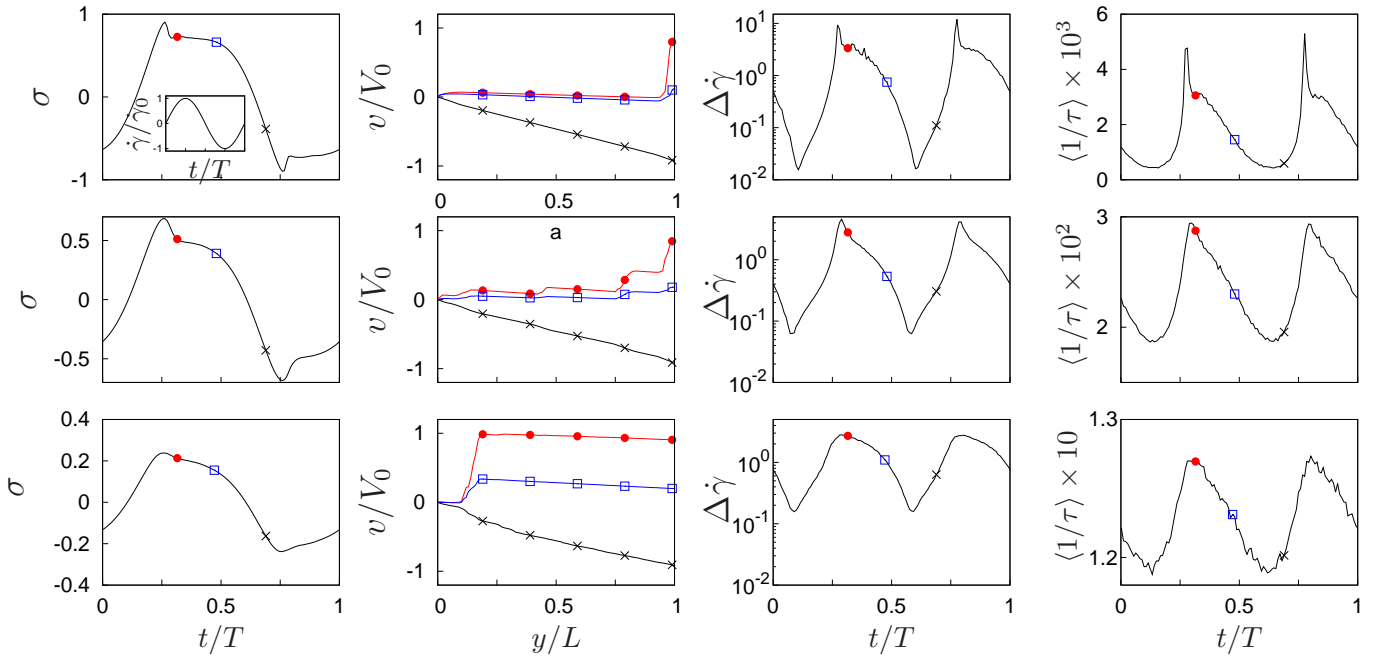}
\caption{Response of the SGR model to LAOStrain of amplitude $\gamma_0=1.59$ and frequency $\omega=0.001$ for three different noise temperatures: $x=0.3$ (top row), $x=0.7$ (middle row) and $x=1.1$ (bottom row).  Sample age before shearing commenced $\tw=10$ for $x=0.3,0.7$, and $\tw\to\infty$ (i.e., sample initialized in  equilibrium) for $x=1.1$. Data shown for cycle number $N=50$. Signals show: (first column) shear stress as a function of time over a cycle, (second column) snapshot shear banded velocity profiles normalized by $V_0=\gdot_0 L$ at three times over a cycle, (third column) inverse effective sample age as a function of time over a cycle, and (fourth column) degree of shear banding as a function of time over a cycle.  Flow profiles in the second column are shown for the times indicated by the corresponding symbols in the other columns.  Number of streamlines $n=100$. Number of SGR elements per streamline $m=100$. Diffusivity $w=0.1$. Toy curvature parameter, $\kappa=0$. Initial heterogeneity $\epsilon=0.1$.  }
\label{fig:velo_prof}
\end{figure*}

As a check of our code, we compared the results of runs with a single streamline $n=1$, for which the flow is homogeneous by definition, with those of analytical calculations for the original homogeneous model~\cite{Sollich1998,Fielding:2000}. We did so for both small and large amplitude oscillatory strain and stress, and for the model's transient and steady state response to a shear startup and an imposed step stress. We do not show data for all these, but a sample comparison is shown in Fig.~\ref{fig:comp_sollich} for large amplitude oscillatory shear strain at a noise temperature $x=1.5$ above the glass point. Ignoring the transient behavior, the stress response to such a deformation can be written
\begin{equation}
\sigma (t) = \gamma_0 [ G' \sin(\omega t) + G'' \cos(\omega t) ] + \delta \sigma(t) \ ,
\label{eqn:moduli}
\end{equation}
where $G'(\omega,\gamma_0)$, and $G''(\omega,\gamma_0)$ are the storage and loss moduli that characterize the response of the system at the level of the fundamental mode, with the all the higher harmonic stress contributions being measured by the residual $q(\omega,\gamma_0)$, where
\be 
\label{eqn:residual}
q^2=\frac{\int dt [\delta \sigma(t)]^2}{\int dt [\sigma(t)]^2}.
\ee
To within numerical noise, we find excellent agreement between these quantities computed within our stochastic simulation and the same quantities computed from analytical expressions.

\section{Reported measures}

In what follows, we shall be interested in the extent to which the response of a sample to large amplitude time-periodic shear protocols is shear banded, for different values of the amplitude and frequency of the imposed oscillation. To characterize the degree of shear banding in the sample at any time $t$, we measure the spatial variance in the shear rate across the flow cell
\be
\label{eqn:degree_banding_time}
\Delta\gdot(t)=\frac{1}{N_0}\sqrt{\langle
  \gdot^2\rangle_i-\langle\gdot\rangle_i^2}.  
\ee
where $\langle\cdots\rangle_i$ denotes an average across streamlines. 
For large amplitude oscillatory shear strain, the normalization factor $N_0=\gdot_0$. For large amplitude square/triangular/sawtooth wave strain rate, $N_0=\omega\gamma_0$. In normalizing in this way by a quantity that scales with the peak strain rate over the cycle as a whole, rather than the strain rate $\gdot(t)$ at the given time $t$, Eqn.~\ref{eqn:degree_banding_time} in fact provides a conservative estimate of the degree of banding (while also reducing the error that can arise due to noise when the instantaneous rate $\gdot(t)$ is used instead). In large amplitude oscillatory shear stress, the normalization factor $N_0$ is defined as the maximum shear rate observed at any point in the cycle. (Therefore, in LAOStress $N_0$ is calculated numerically, whereas in imposed-strain protocols it is known upfront.)

In summarizing the response of the system over a broad range of values of the amplitude and frequency of the imposed flow, we sometimes instead report the degree of banding as defined in Eqn.~\ref{eqn:degree_banding_time}, now averaged over a cycle:
\be
\label{eqn:degree_banding_average}
\cycband=\langle\Delta\gdot(t)\rangle_T,
\ee
where $\langle\cdots\rangle_T$ denotes a time average over a cycle.
Indeed to reduce noise we further average $\cycband$ over the
$N=50-100$th cycles.  Typically, a value $\cycband>0.5$ in this
cycle-averaged measure corresponds to significant banding seen in
visual inspection of the velocity profiles. For large amplitude
oscillatory stress, we report the degree of banding maximized over a
cycle, $\maxband$.

Finally, we shall find it useful to characterize the way in which the
effective age of the sample varies as a function of time over a cycle.
To do this, we define
\be
\label{eqn:age}
\langle 1/\tau \rangle(t) = \sum_{i=1}^{i=n} \sum_{j=1}^{j=m}\exp ( -
(E_{ij} - k l_{ij}^2)/x)/(m n),
\ee
the inverse of which gives a measure of the sample's age.

All our results below are presented for just a single simulation run, apart from in Fig.~\ref{fig:transition}, which averages over twenty five runs.

\section{Results: Large Amplitude Oscillatory Shear Strain}
\label{sec:LAOStrain}

In this section, we report our results for the response of the soft glassy rheology (SGR) model to a large amplitude oscillatory shear strain (LAOStrain). In Fig.~\ref{fig:velo_prof}, a complete cycle of the oscillation is shown for three different values of the noise temperature: two in the glass phase, $x=0.3$ (top row) and $x=0.7$ (middle row), and one just above the glass point, $x=1.1$ (bottom row). The amplitude $\gamma_0$ and frequency $\omega$ of the imposed oscillation is the same in each case.  The origin of time is chosen to be that at which the strain rate switches from negative to positive (inset in the top left panel).  For each noise temperature, we show the stress as a function of time over one cycle (first column), snapshot shear banded profiles at three different times (second column), the degree of shear banding as a function of time over the cycle (third column) and the inverse of the average stress relaxation time, which can be taken as effectively being the inverse sample age, as a function of time (fourth column). The sample age before shearing commenced $\tw=10$ for the noise temperatures $x=0.3,0.7$ in the glass phase in the top two rows, while  $\tw\to\infty$ (corresponding to a sample initialized in equilibrium) for the noise temperature $x=1.1$ above the glass point in the bottom row.

Consider the first half of the cycle, during which the strain rate is positive and the sample is straining in the forward direction.  Initially, when the strain rate has only just switched from negative to positive after the end of the previous cycle, the imposed flow is weak and the sample is old and aging. This can be seen by the fact that the inverse effective sample age (fourth column) as defined by Eqn.~\ref{eqn:age} is initially small and decreasing.  The associated rheological response is accordingly predominantly elastic, with the stress initially increasing approximately linearly with the time and accumulating strain (first column).

As the shear rate progressively increases towards its maximum positive value at the end of the first quarter cycle, the effect of the stronger shear is then to rejuvenate the sample, with $\langle 1/\tau \rangle$ increasing to a maximum.  Associated with this rejuvenation is an overshoot in the stress as a function of time, with the sample then yielding into a flowing regime where the stress remains relatively constant as a function of time. As the shear rate progressively drops towards the end of the first half cycle, the stress likewise drops and the inverse age decreases (ie, the sample ages again).  The same sequence of processes then repeats in reverse, with appropriate changes in sign, during the second half of the cycle in which the strain rate is negative and the sample strains in the reverse direction.

Closely associated with the stress overshoot and subsequent process of yielding during each half of the cycle is the formation of shear bands.  This can be see by the snapshot velocity profiles, $v(y)=\int_{0}^{y} \dot{\gamma}(y') dy'$, in the second column, which deviate strongly from the linear form they would have in the absence of banding. At any time $t$ we take as a measure of the degree of banding the quantity defined in Eqn.~\ref{eqn:degree_banding_time}.  Our results for this quantity as a function of time over the cycle are shown in the third column of Fig.~\ref{fig:velo_prof}.  As can be seen, this measure increases sharply around the time of the stress overshoot, then subsequently decays.

Comparing the three rows in Fig.~\ref{fig:velo_prof}, we find the response is broadly the same in the model's glass phase, where its underlying steady state flow curve has a yield stress, and just above the glass point, where the flow curve is of power law fluid form. The lower noise temperatures however show a more pronounced alternation between aging and rejuvenation within each cycle, and a stronger stress overshoot.  The peak of the degree of banding over a cycle is also slightly stronger for $x<1$.

\begin{figure}
	\centering
	\includegraphics[width=8cm]{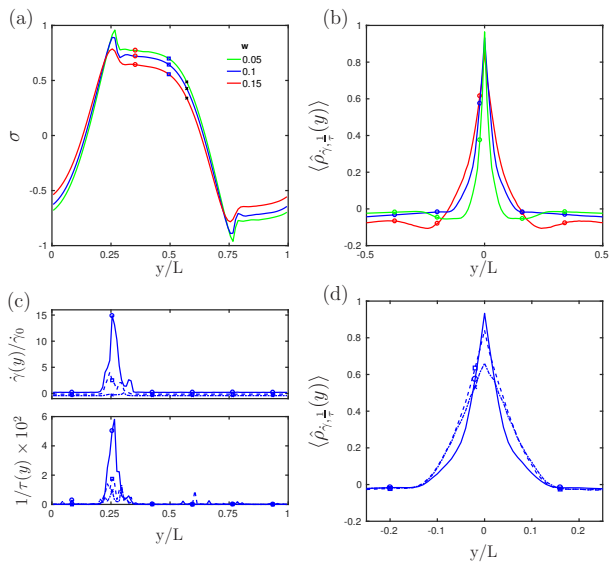}
	\caption{\textbf{(a)} Oscillatory shear stress response of an SGR model with noise temperature $x=0.3$ at $\gamma_0=1.59$, and $\omega=0.001$ for different weighting factors $w=0.05,0.1,0.15$ of stress diffusivity; and \textbf{(b)} the corresponding normalized cross correlation between local shear rate $\gdot(y)$ and inverse age $\tauinv(y)$ at the time indicated by $\circ$ are shown. For $w=0.1$, the \textbf{(c)} shear rate, inverse age profiles, and \textbf{(d)} normalized cross-correlation are shown at different times marked by the symbols in \textbf{(a)}. The other model parameters are n=100, m=100, $t_w=10$. \label{fig:corr} } 
\end{figure}

As seen by comparing the third and fourth columns of Fig.~\ref{fig:velo_prof}, there is a strong temporal correlation between the degree of shear banding and the inverse sample age averaged across the sample. To explore the link between these two quantities in more detail, we now examine the spatial cross-correlation between the local shear rate inside the sample and the local inverse sample age. To do this, we measure the normalised cross-correlation between the local inverse sample age $\tauinv(y)$ and shear rates $\gdot(y)$ at different streamlines as shown in Fig.~\ref{fig:corr}. The discrete cross-correlation function between $\gdot, \tauinv$ between streamlines j apart is defined as 
\begin{equation}
	\rho_{\gdot\tauinv}(j)=
	\begin{cases}	
	\sum_{i=0}^{n-j-1}\left(\gdot(i+j) - \overline{\gdot}\right)\left(\tauinv(i) - \overline{\tauinv}\right),\  j \ge 0\\
	\rho_{\gdot\tauinv}(-j),\  j<0 \ ,
\end{cases}
\end{equation}
where i indicates streamline number, n is the total number of streamlines, and the overline denotes the mean across the sample. The normalised cross-correlation function is given by
\begin{equation}
	\hat{\rho}_{\gdot\tauinv}(j)=\frac{1}{\sqrt{\rho_{\gdot\gdot}(0)\rho_{\tauinv\tauinv}(0)}}\rho_{\gdot\tauinv}(j)\ ,
\end{equation}
where $\rho_{\gdot\gdot}(0)$ is the autocorrelation function for \gdot.  Similar to the stress-signal, this normalised cross-correlation function $\hat{\rho}_{\gdot,\tauinv}$ can then be averaged over multiple cycles to reduce the noise, and expressed as a function of distance $y/L$ rather than the streamline number.  

The normalised cross-correlation function allows us to explore how the spatial correlation between the inverse sample age and local shear rate depends on the weighting factor w for stress diffusivity. From Fig.~\ref{fig:corr} (b), it is clear that width of the cross-correlation $\crcor$ increases with increase in the weighting factor of diffusivity, as is to be expected. The maximum amplitude of the correlation is highest immediately following the stress overshoot, and decreases as the shear rate changes direction as shown in Fig.~\ref{fig:corr} (d), which can also be qualitatively inferred by comparing the \gdot, \tauinv profiles given in Fig.~\ref{fig:corr} (c). Thus, over a LAOS cycle, the average inverse sample age can indicate shear banding, and the the local inverse sample age is correlated with the region of shear banding.

\begin{samepage}
\begin{figure}[!htbp]
\includegraphics[width=8.0cm]{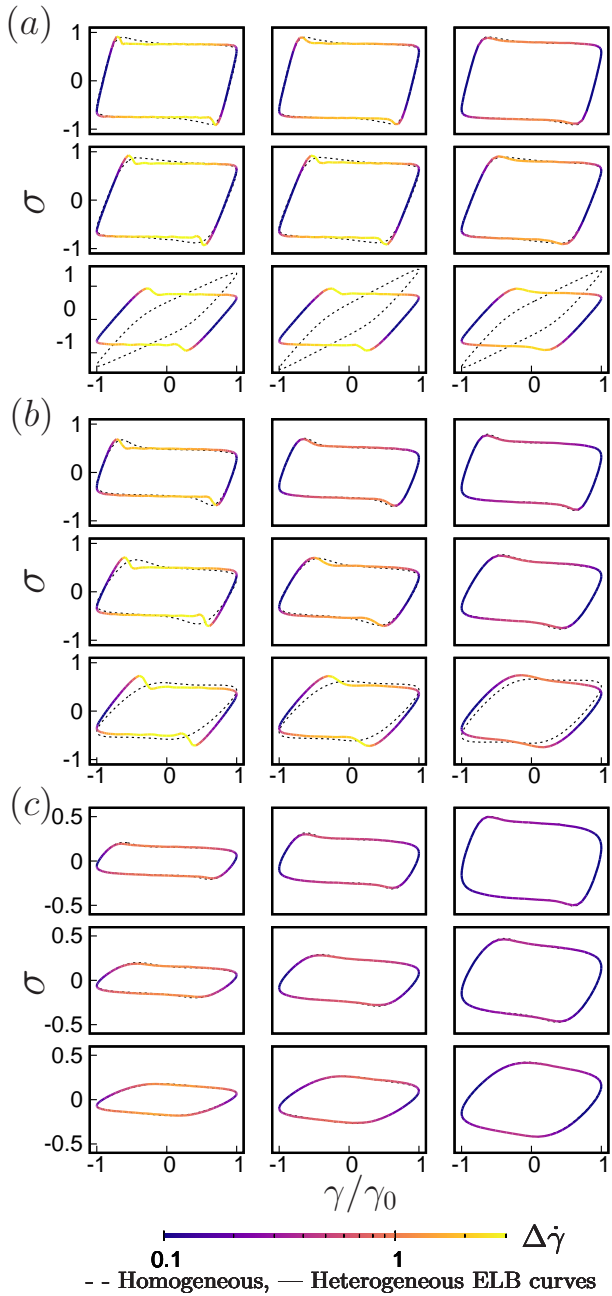}
\caption{Elastic Lissajous-Bowditch curves for the homogeneous (dashed lines) and heterogeneous (solid lines) SGR model in LAOStrain for noise temperatures $x=0.3,0.7,1.1$ in panels (a) to (c) downward.
  Initial sample age before shearing commenced $\tw=10$ in each case.
In the heterogeneous calculations, the instantaneous degree of banding $\Delta \dot{\gamma}$ is indicated by the color-scale. The grid of values of $\gamma_0,\omega$ is the same in each panel, and indicated by crosses in Fig.~\ref{fig:phasediag}.  Data averaged over 50th to 100th cycles.  Heterogeneous calculations have: number of streamlines $n=25$, number of SGR elements per streamline $m=100$, diffusivity $w=0.05$, toy cell curvature $\kappa=0$, and initial heterogeneity $\epsilon=0.1$. Homogeneous calculations have $m=1000$ SGR elements.}
\label{fig:ELB}
\end{figure}

\begin{figure}[!htbp]
\includegraphics[width=8.0cm]{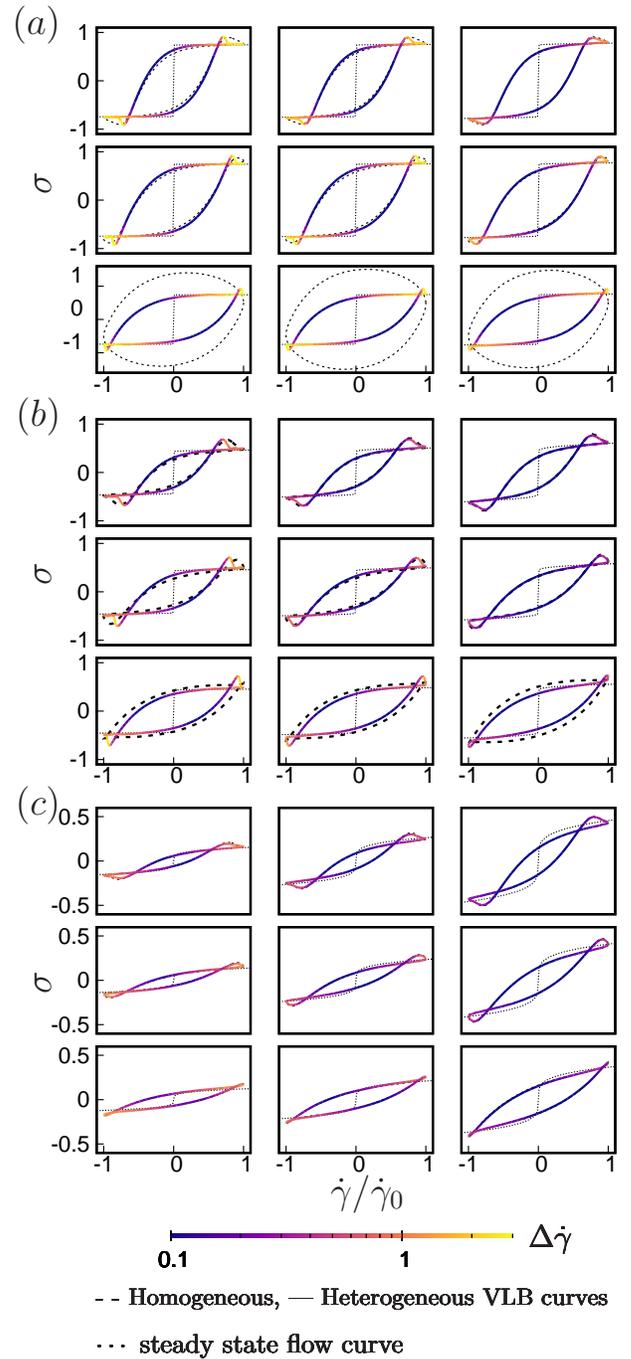}
\caption{Viscous Lissajous-Bowditch curves for the homogeneous (dashed lines) and heterogeneous (solid lines) SGR model in LAOStrain for noise temperatures. Parameter values as in Fig.~\ref{fig:ELB}.  Thin dotted lines show steady state flow curve $\sigma(\gdot)$.  }
\label{fig:VLB}
\end{figure}
\end{samepage}

A common way of visualizing the response of a viscoelastic material to an imposed large amplitude oscillatory shear strain is parametrically to plot the stress as a function of strain over the course of a cycle, to give the so-called elastic Lissajous-Bowditch (ELB) curve; or as a function of strain-rate over the course of cycle to give the viscous Lissajous-Bowditch (VLB) curve~\cite{Ewoldt:2008}.  A grid of such figures plotted for different values of the amplitude $\gamma_0$ and frequency $\omega$ of the imposed oscillation then gives a so-called Pipkin diagram, which is commonly used for rheologically fingerprinting viscoelastic fluids.

Our results for Pipkin diagrams computed in the soft glassy rheology model are shown in Figs.~\ref{fig:ELB} and~\ref{fig:VLB}, in the ELB and VLB representations respectively. In each case we explore the same three noise temperatures as in Fig.~\ref{fig:velo_prof}, although now the initial sample age before shearing commenced $\tw=10$ in each case. The solid lines pertain to the heterogeneous model that takes shear banding into account. The dashed lines are for simulations that impose upfront a purely homogeneous flow, disallowing any possibility of shear banding.

For a simple linear elastic solid, the ELB curve would comprise a straight line through the origin. In contrast, a purely viscous liquid would give an ellipse.  In the SGR model, the ELB curves for low imposed strain amplitudes indeed show purely elastic response. (We do not present these here.)  In contrast, for strain amplitudes $\gamma_0>1$ we see highly nonlinear ELB curves.  Strongly nonlinear ELB curves have been observed experimentally in soft glassy materials in Refs.~\cite{Renou2010,Rogers:2011,Poulos2015}.

These ELB curves contain essentially the same information as discussed in the context of Fig.~\ref{fig:velo_prof} above, but with time now as a hidden parameter that increases as the curve is explored in the clockwise direction during the course of any LAOS cycle.  The bottom-left to top-right sector corresponds to the positive strain-rate half of the cycle, in which the sample is straining in the forward direction. With this in mind, we now identify in the ELB curves a sequence of physical processes~\cite{Rogers:2011} corresponding to the alternating competition over the course of each cycle between glassy aging and flow rejuvenation, and between elastic and viscous response.

The bottom-left of the ELB curve corresponds to the time at which the strain-rate switches from negative to positive and the sample starts being sheared in the forward direction. Initially this shear is of low rate and the sample accordingly is old and aging (for the low frequencies $\omega <1$ to which the SGR model applies), with rather elastic rheological response: the stress initially increases linearly with strain.  As the shear rate then progressively increases during the first quarter cycle, the increasingly strong shearing acts to rejuvenate the sample. We then see a yielding process in which the stress goes through an overshoot as a function of strain, before declining to a flowing regime in which it remains almost constant.  The same sequence of processes then repeats in reverse, during the negative straining half cycle, clockwise from top right to bottom left in the ELB curve.

\begin{figure*}[!tbp]
\includegraphics[width=16.0cm]{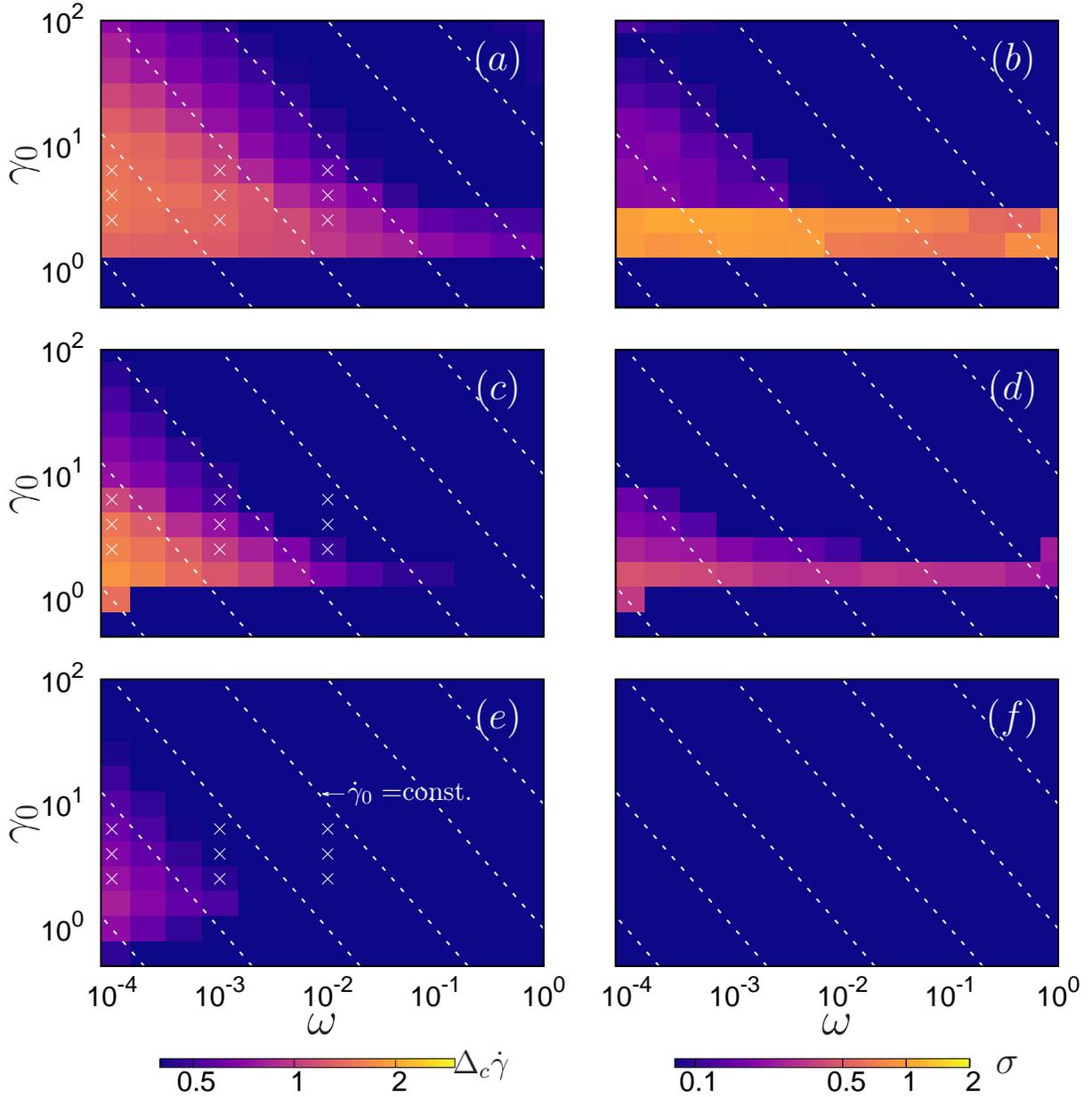}
\caption{Left panels (a,c,d): dynamic phase diagrams showing the
  cycle-averaged degree of banding in the heterogeneous form of the
  soft glassy rheology model in large amplitude oscillatory shear for
  $x=0.3,0.7,1.1$ respectively. Dashed lines show constant
  $\dot{\gamma}_0$.  The $\times$ indicate the grid of
  $\gamma_0,\omega$ values explored in more detail in the ELB and VLB
  curves of Figs.~\ref{fig:ELB} and~\ref{fig:VLB}. Initial sample age $\tw=10.0$ for all
  three noise temperatures. Data averaged over 50th to 100th
  cycles. Right panels (b,d,f) show
  counterpart discrepancy between the ELB curves calculated within the
  assumption of homogeneous flow, and those calculated allowing for
  shear banding, for the same parameters. Number of streamlines
  $n=25$, number of SGR elements per streamline $m=100$, diffusivity
  $w=0.05$, toy cell curvature $\kappa=0$, and initial heterogeneity
  $\epsilon=0.1$.   }
\label{fig:phasediag}
\end{figure*}

In each of the ELB curves, the colorscale shows the degree of banding $\Delta \gdot$ at any point in the cycle. Consistent with our discussion of Fig.~\ref{fig:velo_prof} above, we find the appearance of shear banding to be closely associated with an overshoot in the signal of stress as a function of strain in the ELB curves.  Typically, shear bands form as the overshoot is approached and persist for some time as the stress declines afterwards.  This behavior is strongly reminiscent of transient shear banding associated with stress overshoot in the startup of shear of a constant rate~\cite{Moorcroft2011}, as summarized in Sec.~\ref{sec:introduction} above.

For an ergodic viscoelastic fluid with a fixed characteristic stress relaxation time $\tau$, we expect a sequence of LAOS experiments repeated with the same amplitude $\gamma_0$ for progressively lower values of the imposed frequency $\omega$ to reveal a progression from elastic-like response in the high frequency regime $\omega\tau\gg 1$ to viscous-like response in the low frequency regime $\omega\tau\ll 1$.  Furthermore, in the limit $\omega\to 0$ we expect to recover a scenario in which the fluid repeatedly sweeps quasistatically up and down its viscous steady state flow curve as the strain rate slowly increases and decreases over the course of each cycle.  A Lissajous Bowditch plotted in the viscous representation of stress as a function of strain rate should then correspond to the fluid's underlying steady state flow curve.  For any material in which the constitutive curve is purely monotonic, shear banding would be impossible in this quasi-static limit. Such a scenario was indeed explored in ergodic polymeric fluids in Refs.~\cite{Adams2009,Carter2015}.

However, in the glass phase $x<1$ of the soft glassy rheology model we find no such progression with decreasing frequency leftwards along any row of the Pipkin grids in Figs.~\ref{fig:ELB}a,b) and~\ref{fig:VLB}a,b).  Even at the lowest accessible frequencies we still observe strongly elastic response, in some part of the cycle at least, with the stress increasing almost linearly with strain in the ELB representation $\sigma(\gamma)$. In the viscous representation $\sigma(\gdot)$, we never find the VLB curve to approach the underlying steady state flow curve: instead, it displays markedly open loops even at the lowest frequencies accessible numerically. Furthermore, we find that strong shear banding likewise persists, despite the underlying constitutive curve being monotonic.

This highly counterintuitive behavior arises from a basic competition within each cycle between glassy aging in the low shear rate phase of the cycle alternating with flow-induced rejuvenation, yielding and the associated shear banding in the high shear rate part of the cycle.  Put simply: an aging material has no fixed characteristic relaxation rate $1/\tau$ against which we can set the frequency $\omega$ of the imposed oscillation to be small.  This finding has far reaching implications for the flow of aging soft glasses, suggesting a strong predisposition to shear banding even in imposed flow protocols of arbitrarily slow time-variation~\cite{SM}.

In contrast, for noise temperatures $x>1$ above the model's glass point, the underlying flow curve is of power law fluid form. In the absence of flow, true aging is absent~\cite{Fielding:2000}, although very long transients associated with sluggish relaxation timescales may nonetheless still arise. In consequence, in a sequence of LAOS experiments performed at fixed oscillation amplitude $\gamma_0$ for progressively smaller values of the imposed frequency $\omega$, the ELB and VLB curves enclose a progressively smaller area. For noise temperatures far enough above the glass point and low enough frequencies, the VLB curves eventually tend to the steady state flow curve, with no associated shear banding.  However, for the noise temperature $x=1.1$ considered here, only just above the glass point, we have not been able to access low enough frequencies to see a return to purely homogeneous response. It would be interesting in future work to explore the response in the low frequency limit for noise temperature just above the glass point.

In Figs.~\ref{fig:ELB} and~\ref{fig:VLB}, we have discussed the response of the SGR model to a series of LAOStrain experiments with a set of imposed amplitude and frequency values $(\gamma_0,\omega)$ arranged on a 3x3 grid. To explore more fully the regimes of amplitude and frequency in which significant banding arises, we show in the left panels (a,c,e) of Fig.~\ref{fig:phasediag} full dynamic phase diagrams, respectively for each of the three noise temperatures $x=0.3,0.7,1.1$. In any such phase diagram, each coordinate pair $(\gamma_0,\omega)$ corresponds to a LAOStrain experiment performed with those given $(\gamma_0,\omega)$.  Represented by the colorscale at each $(\gamma_0,\omega)$ is then the cycle-averaged degree of banding $\Delta_c \gdot$, as defined in Eqn.~\ref{eqn:degree_banding_average}, arising in a LAOS experiment performed with that given strain amplitude and frequency.  We have checked that a value $\Delta_c\gdot > 0.5$ corresponds to strongly visually apparent banding in the flow profiles.

\begin{figure}[!tbp]
\includegraphics[width=7.5cm]{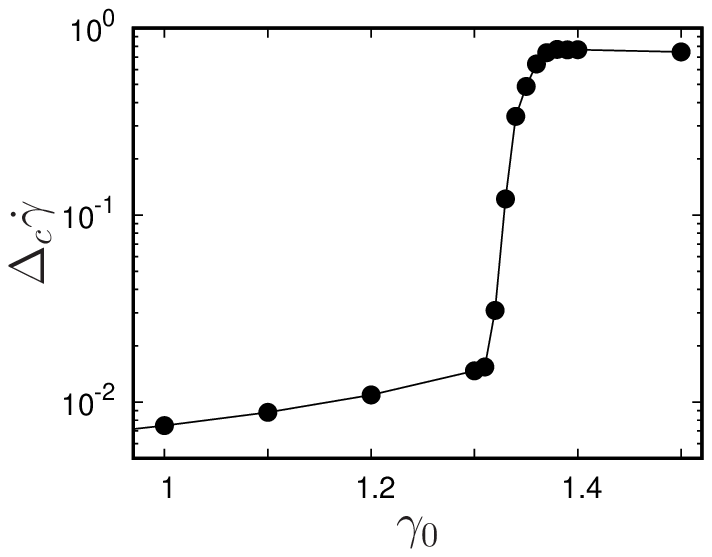} 
\caption{Transition from non-banded to banded flow in the soft glassy rheology model at a noise temperature $x=0.3$ in a series of LAOS experiments performed at a fixed frequency $\omega=0.1$ for increasing values of the strain amplitude $\gamma_0$. Device curvature $\kappa=0.01$, sample age $\tw=10.0$, $\epsilon=0.1$, $w=0.05$. Data averaged over 50th to 100th cycles, and over 25 separate simulation runs. $m=1600$, $n=75$.} 
\label{fig:transition}
\end{figure}

For all the noise temperatures shown, both in the glass phase and just above the glass point, we find significant banding across a significant region of the plane of imposed strain amplitude and frequency: roughly, in the glass phase $x<1$, for strain amplitudes $\gamma_0> 1$ and strain rate amplitudes $\gdot_0=\gamma_0\omega < \gdot_{0{\rm c}}(x)$. (Lines of constant strain rate are shown by the dashed lines in Fig.~\ref{fig:phasediag}.) The value $\gdot_{0{\rm c}}(x)$ of the strain rate amplitude below which significant banding is observed clearly decreases with increasing noise temperature $x$.  Accordingly, the degree of banding for a given pair of values of imposed oscillation and frequency $\gamma_0,\omega$ decreases with increasing $x$. This can be understood by appreciating that for increasing values of $x$ in the model's glass phase, we see less pronounced aging.  Indeed, true aging is eliminated in favor of long transient evolution to a sluggish steady state for $x>1$.  Accordingly, the repeated aging and rejuvenation that underpins the triggering of shear banding in each cycle becomes less pronounced with increasing $x$.

Inspecting again the color maps of the degree of banding as a function of imposed strain amplitude and frequency in the phase diagrams of Fig.~\ref{fig:phasediag}, we see that (at any noise temperature $x$), the transition from non-banded to banded flow, in a series of LAOS experiments performed at a fixed value of the frequency $\omega$ and progressively increasing amplitude $\gamma_0$, appears to be rather sharp. This transition is investigated in Fig.~\ref{fig:transition}, where we indeed see a rather sharp transition to banding with increasing strain amplitude. 

Most theoretical studies of LAOS to date have imposed upfront a homogeneous shear flow, discarding any possibility of shear banding.  However, our results in Figs.~\ref{fig:ELB} and~\ref{fig:VLB} show the danger of calculating rheological fingerprints (ELB or VLB curves) within any such assumption. In each panel of Figs.~\ref{fig:ELB} and~\ref{fig:VLB}, the solid line shows the Lissajous-Bowditch curve in a calculation that properly allows for banding, while the dashed line shows the corresponding curve in a calculation that disallows banding and imposes homogeneous flow. As can be seen, the presence of shear banding can cause a strong discrepancy between these two curves, particularly for strain amplitudes that are only just in the nonlinear regime.

To explore this discrepancy further, in the right panels (b,d,f) of Fig.~\ref{fig:phasediag} we show as a color map in the plane of imposed strain amplitude and frequency the maximum difference in stress $\Delta_m \sigma$ between the homogeneous and heterogeneous calculations. For numerical convenience, this is measured over a time interval $T/10$ following the peak in the stress signal for the heterogeneous flow, where $T$ is the time-period of the oscillation.  (This is indeed the time-interval when any difference between the two signals is most pronounced.)  As can be seen, for the noise temperatures $x=0.3,0.7$ in the glass phase, a strong discrepancy between the homogeneous and heterogeneous calculations is observed for imposed strain amplitudes just into the nonlinear regime $\gamma_0\gtrsim 1$.  For the noise temperature $x=1.1$ above the glass point, where the model shows ergodic power law fluid behavior, this discrepancy is essentially non existent. (However, strong discrepancies were reported in a model of ergodic polymeric fluids in Ref.~\cite{Carter2015}.) An important message of this work is therefore to counsel caution in seeking to fingerprint complex fluids via theoretical calculations that assume homogeneous flow.

\begin{figure*}[!tbp]
\includegraphics[width=15.0cm]{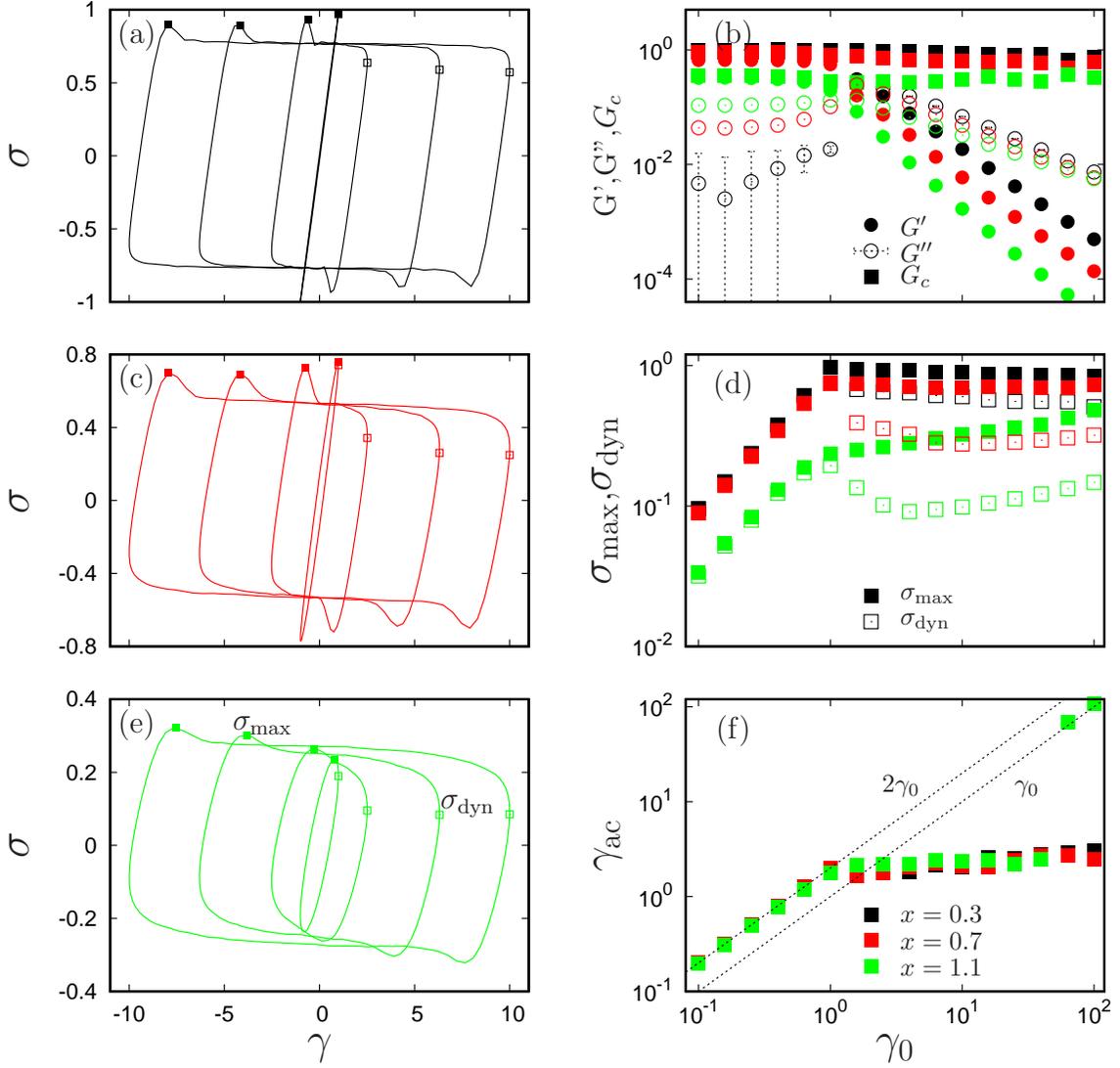}
\caption{(a,c,e) Elastic Lissajous curves of the SGR model in LAOStrain at a noise temperature $x=0.3,0.7,1.1$ respectively. In each case the oscillation frequency $\omega=10^{-3}$, with curves shown for values of the strain amplitude $\gamma_0=1,2.51,6.31, 10$.  (b) The cage modulus $\Gc$ ($\blacksquare$), storage modulus $G'$ ($\bullet$) and loss modulus $G''$ ($\circ$) extracted from a family curves, as a function of imposed strain amplitude for the same frequency $\omega=10^{-3}$. (d) Maximum stress $\sigmamax$ ($\blacksquare$) and dynamic yield stress $\sigmadyn$ ($\square$).  (f) Strain acquired at the stress maxima since strain reversal $\gammaac$ ($\blacksquare$) as defined in the main text. Lower and upper dotted lines in (f) show $\gammaac=\gamma_0$ and $\gammaac=2\gamma_0$ respectively. Initial sample age $\tw=10.0$. Data averaged over 50th to 100th cycles. Number of streamlines $n=25$, number of SGR elements per streamline $m=100$, diffusivity $w=0.05$, toy curvature parameter $\kappa=0$, initial heterogeneity $\epsilon=0.1$. In each of (b,d,f), the color coding with respect to noise temperature matches that of (a,c,e).}
\label{fig:osc_rogers}
\end{figure*}

Finally in this section on large amplitude oscillatory shear strain, we seek to interpret the elastic Lissajous-Bowditch (ELB) curves of the heterogeneous soft glassy rheology model within the framework of a `sequence of physical processes', as introduced by Rogers et al. in Ref.~\cite{Rogers:2011} and applied to yield stress and power law fluids in Ref.~\cite{Rogers:2012}. In particular, we shall compute the various nonlinear quantities proposed by Rogers et al. as being useful measures of the response of yielding materials in LAOStrain.  With this in mind, in the left panels of Fig.~\ref{fig:osc_rogers} we show again ELB curves for our three different noise temperatures $x=0.3,0.7,1.1$, respectively in panels from top to bottom. In each case, we show results for a fixed value of the cycle frequency $\omega=10^{-3}$, for several different values of the imposed strain amplitude $\gamma_0$.

For each such curve we then computed the storage and loss moduli, $G'$ and $G''$, as defined in Eqn.~\ref{eqn:moduli}. These are plotted as a function of the imposed strain amplitude $\gamma_0$ in the top right panel of Fig.~\ref{fig:osc_rogers}, by the filled and open circles respectively. The elastic modulus decreases with increasing $\gamma_0$: initially gently in the linear regime $\gamma_0\lesssim 1$, then much more rapidly in the nonlinear regime $\gamma_0 \gtrsim 1$. The loss modulus $G''$ instead initially increases with $\gamma_0$ in the linear regime, before showing a peak then subsequently decreasing in the nonlinear regime. These forms are consistent with the earlier results of Ref.~\cite{Sollich1998}. In the linear regime, $G'>G''$, with the reverse true in the strongly nonlinear regime. Both quantities decrease with increasing noise temperature $x$, for all values of the imposed strain amplitude $\gamma_0$. In the nonlinear regime, all the quantities shown in Fig.~\ref{fig:osc_rogers} are in a state of cycle-to-cycle invariance (to excellent approximation)~\footnote{See the supplementary material of Ref.\cite{Ranga:2016}}.  In the linear regime, the values of $G'$ and $G''$ slowly age.  This was studied previously~\cite{Fielding:2000} and we do not consider it further here.

The storage modulus $G'$ is intended to characterize the material's elastic response. As just noted, it decreases dramatically through the nonlinear regime to become small at high values of the imposed strain amplitude $\gamma_0$. While this may be a reasonable representation of the response of the material integrated over an entire cycle, $G'$ nonetheless fails to capture the obvious region of elastic response that persists even at large imposed strain amplitudes, in the part of the ELB curves near flow reversal at $\gamma(t)=\pm\gamma_0$, where the stress $\sigma(t)$ is small. Recall the steeply sloping sections of the ELB curves in Fig.~\ref{fig:osc_rogers}a).

To characterize this regime of elastic response near flow reversal,
Rogers et al. defined the `cage modulus':
\be
\label{eqn:cage}
\Gc={\frac{d \sigma}{d \gamma}}\at[\Big]{\sigma=0}.
\ee
Our results for this quantity, extracted from the ELB curves of Fig.~\ref{fig:osc_rogers}a),c,e), are shown in Fig~\ref{fig:osc_rogers}b). In the linear viscoelastic regime $\gamma_0\to0$, it was proved analytically in Ref.~\cite{Rogers:2011} for these yielding materials that $\Gc=G'+G''^2/G'$. We have verified that this relation is indeed satisfied for our data.  Beyond the linear regime, the cage modulus remains almost constant across the full range of $\gamma_0$ considered, even as the storage modulus falls dramatically at large $\gamma_0$. In this way, the cage modulus is able to capture the intra-cycle elasticity observed for small stresses near strain reversal in the ELB curves, even at large values of the imposed strain amplitude $\gamma_0$.  At any given imposed $\gamma_0$, the cage modulus $\Gc$ decreases with increasing noise temperature $x$. 

\begin{figure*}[!tbp]
\includegraphics[width=15.0cm]{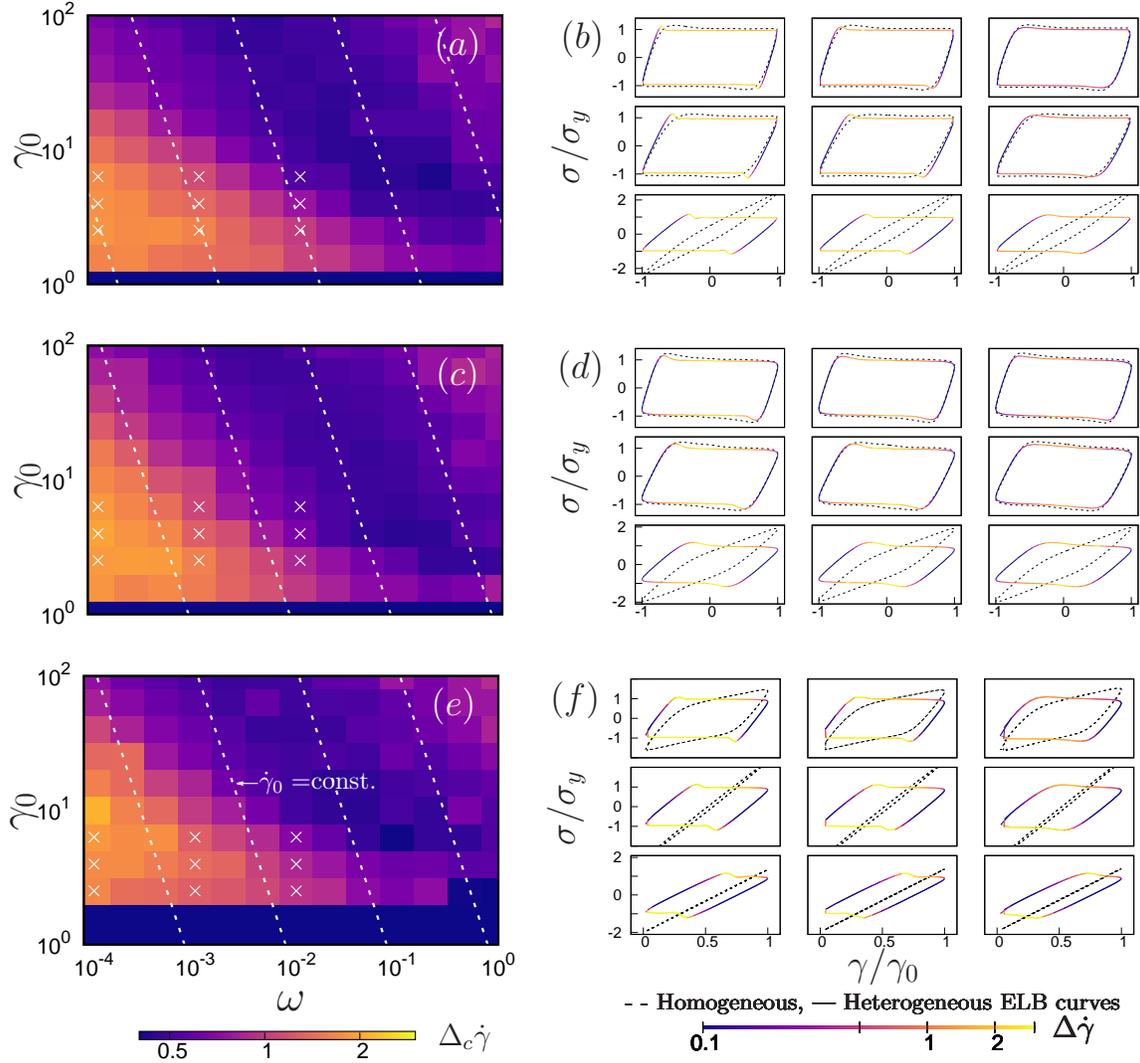}
\caption{Left panels (a,c,e): dynamic phase diagrams showing the cycle-averaged degree of shear banding in the heterogeneous form of the soft glassy rheology model in large amplitude square, triangle and sawtooth strain rate respectively.  Dashed lines show constant $\dot{\gamma}_0$.  Right panels (b,d,f) show counterpart elastic Lissajous-Bowditch curves for the homogeneous (dashed lines), and heterogeneous (solid lines) models for the grid of $\gamma_0,\omega$ values indicated by $\times$ in the left panels.  In the heterogeneous calculations, the instantaneous degree of banding $\Delta \dot{\gamma}$ is indicated by the color-scale.  Noise temperature $x=0.3$. Initial sample age $\tw=10.0$.  Data averaged over 50th to 100th cycles.  Heterogeneous runs have: number of streamlines $n=100$, number of SGR elements per streamline $m=100$, diffusivity $w=0.1$, initial heterogeneity $\epsilon=0.1$, and toy curvature parameter $\kappa=0$.  Homogeneous runs have  $m=1000$ SGR elements.}
\label{fig:dif_prot}
\end{figure*}

Another measure that is commonly discussed in relation to yield stress fluids is that of the `yield stress' itself. Indeed several different quantitative definitions are commonly used to characterize this intuitive concept~\cite{Dinkgreve:2016}. Broadly, the stress above which the material starts flowing is termed the static yield stress, while the stress below which it stops flowing is called the dynamic yield stress.  In the SGR model, the maximum stress that can be maintained indefinitely without the material flowing with a non-zero strain rate at long times (the `static yield stress'), and the minimum stress obtained in sweeping the imposed strain rate $\gdot\to 0$ (the `dynamic yield stress') are the same, and give a well defined `yield stress' $\sigmay(x)$ that is non-zero for $x<1$~\cite{Fielding:2000}.

In this context of oscillatory flows, Rogers et al.~\cite{Rogers:2011} sought to obtain measures of the yield stress from the ELB curves. In particular, they defined the static yield stress to be maximum stress $\sigmamax$ in the ELB curve, and the dynamic yield stress $\sigmadyn$ to be the value of the stress at the point where the strain is maximum, $\gamma(t)=\gamma_0$. We have marked these quantities on the ELB curves of Fig.~\ref{fig:osc_rogers}a,c,e) by filled and open squares respectively. Fig.~\ref{fig:osc_rogers}d) plots the same quantities (with the same symbol key) as a function of imposed strain amplitude $\gamma_0$. In the linear viscoelastic regime $\gamma_0\to 0$ the two quantities coincide and follow a linear elastic increase with $\gamma_0$.  In the nonlinear regime $\gamma_0\gtrsim 1$, they start to separate, with the dynamic quantity $\sigmadyn$ becoming lower than the static one $\sigmamax$.  At any fixed $\gamma_0$, both $\sigmamax$ and $\sigmadyn$ decrease with increasing noise temperature $x$, as expected.  However there is a clear difference between the dependence of the static yield stress $\sigmamax$ on the imposed strain amplitude $\gamma_0$ in the nonlinear regime $\gamma_0\gtrsim 1$ for noise temperatures in the glass phase and those above the glass point. In the glass phase, it is roughly constant. Above the glass point, it increases with increasing $\gamma_0$.

Another measure commonly discussed for yield stress fluids is that of the yield strain. Several different definitions again exist.  In the present context of LAOStrain we consider $\gammaac$, defined as the strain acquired between the point of strain reversal (where $\gamma=-\gamma_0$) and the point of absolute maximum stress in the cycle following the strain reversal (i.e., the point shown by the filled squares in Fig.~\ref{fig:osc_rogers}a).  Our results for this quantity are shown in Fig.~\ref{fig:osc_rogers}(f), with solid squares. In the linear viscoelastic regime, the ELB curve is a straight line through the origin, giving $\gammaac=2 \gamma_0$.

The trends reported in the SGR model in Fig.~\ref{fig:osc_rogers}d)
for $\sigmamax$ and $\sigmadyn$ and in Fig.~\ref{fig:osc_rogers}f) for
$\gammaac$ broadly resemble those reported experimentally in star
polymers~\cite{Rogers:2011}, a hard sphere
suspension~\cite{vanderVaart:2013}, and a colloidal gel
~\cite{Kim:2014}, though we do not attempt quantitative comparison.

\section{Results: Large Amplitude Square/Triangle/Sawtooth Wave strain  rate}
\label{sec:LAOSothers}

In the previous section, we presented the results of theoretical calculations suggesting that soft glassy materials exhibit shear banding in large amplitude oscillatory shear strain, across a broad range of values of the amplitude $\gamma_0$ and frequency $\omega$ of the imposed oscillation. In the glass phase, we showed that this effect persists even at the lowest frequencies accessible numerically, even though the model's underlying constitutive curve is purely monotonic, rendering it incapable of supporting shear bands as the true steady state response to a steadily imposed shear of constant rate. We interpreted this counterintuitive behavior as arising from an alternating competition within each cycle between glassy aging in the low strain rate phase, and flow-rejuvenation in the high strain rate phase.

In this section, we show that same scenario also arises in other large amplitude time-periodic shear strain protocols. While being far from conclusive (we perform our calculations in just one particular model of soft glasses, for four different strain-imposed waveforms), this finding has potentially far reaching implications for the rheology of soft glasses more generally, in suggesting a rather generic predisposition to shear banding in time-varying flows of any waveform, even in the limit of an arbitrarily slow time-variation.

With these remarks in mind, we consider now the protocols of large amplitude square, triangle and sawtooth wave strain rate, as sketched in Fig.~\ref{fig:protocols}b)-d). (These imposed flows are in fact the basis functions for examining the oscillatory shear stress response of materials as proposed by Klein et al.~\cite{Klein:2007}.) Corresponding to the dynamic phase diagram of the cycle averaged degree of shear banding $\cycband$ shown in Fig.~\ref{fig:phasediag} for oscillatory shear flow, the counterpart phase diagrams for these other three protocols are shown in the left panels of Fig.~\ref{fig:dif_prot}, for a single noise temperature in the model's glass phase. We indeed observe significant banding for a large range of values of the amplitude $\gamma_0$ and frequency $\omega$ of the imposed oscillation, for all three protocols.  Perhaps surprisingly, even the quantitative degree of banding is similar in each case, and is seen over a similar region of the $\gamma_0,\omega$ plane, though with slightly less banding in the sawtooth case.

In the right panels (b,d,f) of Fig.~\ref{fig:dif_prot}, we show ELB
curves corresponding to the grid of $\gamma_0,\omega$ values indicated
in the counterpart phase diagrams in panels (a,c,e).  In each case, we
find a sequence of physical processes similar to that described above
for LAOStrain. The local degree of banding $\Delta \gdot(t)$ is
indicated by the color scale round each cycle. As can be seen, the
onset of banding is again closely associated with the stress overshoot
in each case, closely reminiscent of banding associated with stress
overshoot in the simpler protocol of shear startup~\cite{Moorcroft2011}.
As in LAOStrain, we see a significant difference between the ELB
curves as calculated allowing shear bands to form (solid lines) and
those for purely homogeneous shear (dashed lines), particularly for
imposed strain amplitudes in the region of transition from no banding
to banding.

\begin{figure}[!tbp]
\includegraphics{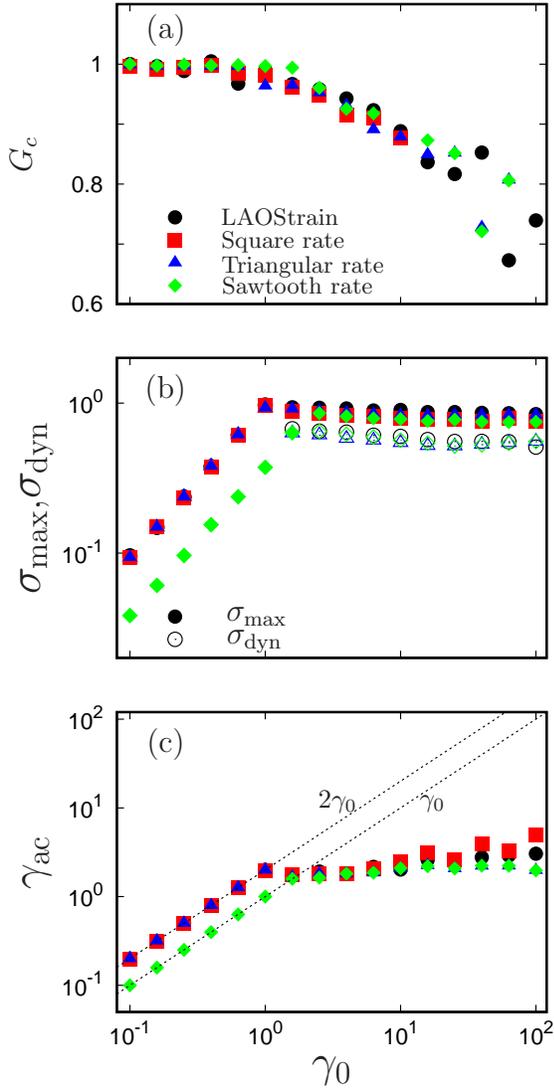}
\caption{ (a) Cage modulus $\Gc$, (b) maximum stress $\sigmamax$ (filled symbols) and dynamic yield stress $\sigmadyn$ (open symbols) and (c) strain acquired between the point of strain reversal and that of maximum stress. In each case data are shown for LAOStrain ($\bullet$), square wave strain rate (\textcolor{red}{$\blacksquare$}), triangular wave strain rate (\textcolor{blue}{$\blacktriangle$}) and sawtooth strain rate (\textcolor{green}{$\blacklozenge$}). Noise temperature $x=0.3$. Initial sample age $\tw=10.0$. Data averaged over 50th to 100th cycles. The lower and upper dotted lines in (c) show $\gammaac=\gamma_0$ and $\gammaac=2\gamma_0$ respectively. Number of streamlines $n=100$, number of SGR elements per streamline $m=100$, initial heterogeneity $\epsilon=0.1$, toy cell curvature $\kappa=0$, diffusivity $w=0.1$.}
\label{fig:prot_rogers}
\end{figure}

To characterise in more detail the ELB curves of these three alternative protocols, we discuss finally the non-linear measures discussed in the context of LAOStrain in Sec.~\ref{sec:LAOStrain}.  As seen in panel (a) of Fig.~\ref{fig:prot_rogers}, the cage modulus is approximately the same for all the four protocols.  The maximum stress $\sigmamax$ and the stress $\sigmadyn$ at the point of flow reversal $\gamma=\gamma_0$ are shown in panel b).  The strain $\gammaac$ acquired between the point of flow reversal and the point at which the stress attains its maximum value is shown in panel c).  For all four protocols, in the linear regime we see essentially elastic response in which each of $\sigmamax$, $\sigmadyn$ and $\gammaac$ increases linearly with the strain amplitude $\gamma_0$ (though with a lower prefactor for $\sigmadyn$ and $\gammaac$ in the case of the sawtooth wave because its imposed strain range is $0$ to $\gamma_0$, compared with $-\gamma_0$ to $\gamma_0$ for the other three protocols).

\section{Results: Large Amplitude Oscillatory Stress}
\label{sec:LAOStress}

\begin{figure}[!tbp]
\includegraphics[width=8cm]{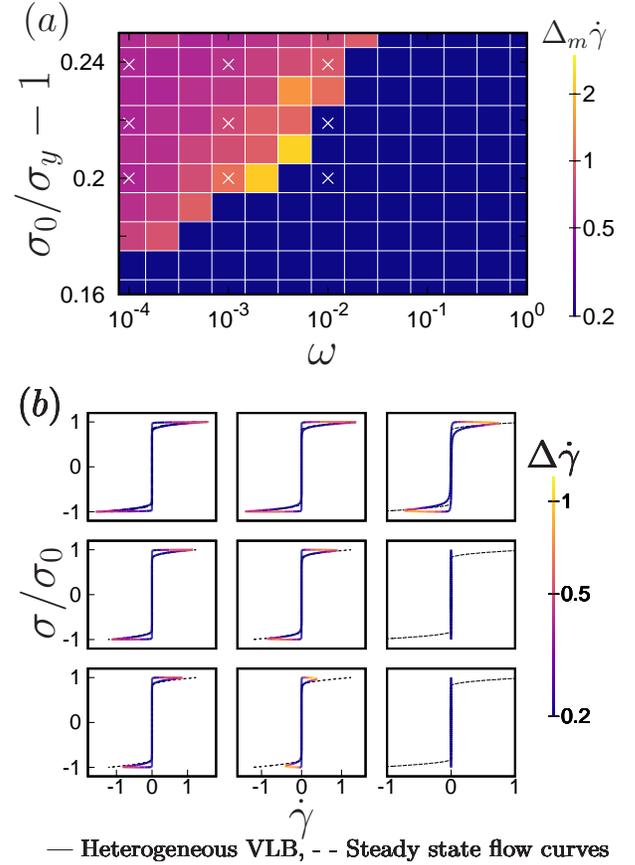}
\caption{\textbf{Top:} Dynamic
  phase diagram showing shear banding in oscillatory shear stress
  protocol for the SGR model with a noise temperature $x=0.3$. \textbf{Bottom:} Viscous Lissajous
  Bowditch curves corresponding to $\times$ symbols in the top panel,
  with the degree of banding shown by the color scale. Initial sample
  age $\tw=10.0$, data averaged over 50th to 100th cycles. Thin dotted
  lines show steady state flow curves $\sigma(\gdot)$. Number of
  streamlines $n=25$, number of SGR elements per streamline $m=100$,
  diffusivity $w=0.05$, initial heterogeneity $\epsilon=0.1$, toy cell
  curvature $\kappa=0$.}
\label{fig:osc_stress}
\end{figure}

\begin{figure}[!tbp]
\includegraphics[width=8cm]{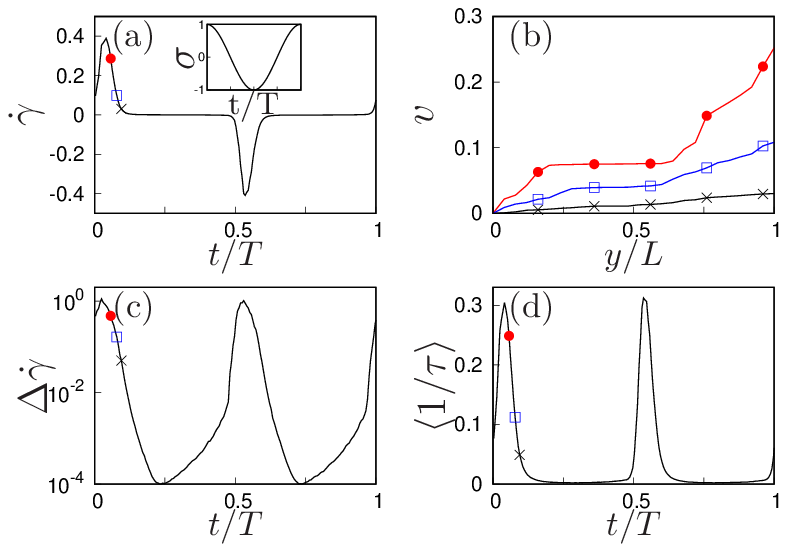}
\caption{Response of the SGR model to LAOStress of amplitude $\sigma_0/\sigma_y-1=0.2$, frequency $\omega=0.001$ for cycle number $N=50$. Signals show (a) shear rate,  (c) degree of shear banding, and (d) inverse effective sample age  as a function of time over a cycle. (b) Snapshots of shear banded velocity profiles corresponding to the symbols in the other figures are shown. Initial sample age $\tw=10$, number of streamlines $n=25$, number of SGR elements per streamline $m=100$, diffusivity $w=0.05$, initial heterogeneity $\epsilon=0.1$, toy curvature parameter $\kappa=0$.}
\label{fig:laostress_velo}
\end{figure}

As summarised in Sec.~\ref{sec:introduction} above, when an initially
well rested sample of soft glass of some age $t=\tw$ is subject to the
switch-on of a stress that is held constant thereafter, it initially
shows a regime of slow creep in which the strain rate progressively
reduces over time.  For imposed stresses above the yield stress, this
regime of slow creep is then followed by a yielding process in which
the strain rate increases towards its final flowing state on the flow
curve. During the time interval in this yielding process over which
the strain rate signal simultaneously curves and slopes upwards as a
function of time, the sample is predicted to be unstable to the
formation of shear bands~\cite{Moorcroft2013a}.

Intuitively, we might expect a large amplitude oscillatory shear stress (LAOStress) protocol loosely to correspond to a repeating sequence of positive and negative step stresses. If a yielding process arises following each of these  steps in each half cycle, we might then intuitively expect to see the formation of shear bands associated with that yielding, by analogy with the banding seen in the simpler step stress protocol just described. With this in mind, we now consider finally the response of the soft glassy rheology model in LAOStress, in its glass phase $x<1$.

In Fig.~\ref{fig:osc_stress} (top) we plot as a color-scale the degree of shear banding maximized over a cycle for a wide range of LAOStress experiments of imposed stress amplitude $\sigma_0$ and frequency $\omega$.  As can be seen, significant shear banding arises across a broad region of the plane of $\sigma_0,\omega$. Banding  persist even at the lowest frequencies accessible numerically (in a manner apparently consistent with it persisting to the limit of zero frequency $\omega\to 0$, were this accessible numerically), as in the strain-imposed protocols considered in previous sections, despite the model's underlying flow curve being purely monotonic, precluding banding as the true steady state response to a constant imposed shear stress.

In the lower panel, we present the corresponding VLB curves for the grid of values of imposed stress amplitude $\sigma_0$ and frequency $\omega$ marked by crosses in the top panel. The time-dependent degree of shear banding is shown as a color-scale round each cycle. The results can be understood as follows. For most of the cycle the stress is below the yield stress, and the shear rate is accordingly small. Once the stress exceeds the yield stress, the sample yields and starts to flow (at low frequencies at least - at higher frequencies there is insufficient time for this to occur). Associated with this yielding process is indeed the formation of shear bands, as predicted by our intuitive argument above.  Noting that the shear banding only arises in a relatively small portion of the cycle in LAOStress, we chose in our color-map in the left panels to show the degree of banding maximized over a cycle.

The response of the system as a function of time round a cycle is shown in more detail in Fig.~\ref{fig:laostress_velo}. Consistent with the preceding discussion, shear bands form in time-regimes where the stress exceeds the yield stress, and the material rejuvenates and starts to flow.

\section{Conclusions}
\label{sec:conclusions}

In this work, we have studied in detail the response of soft glassy materials, including both yield stress fluids and power law fluids, to large amplitude time-periodic flow protocols, in the context of the soft glassy rheology model. For each of large amplitude oscillatory shear strain, large amplitude square wave strain rate, large amplitude triangular wave strain rate, large amplitude sawtooth strain rate and large amplitude oscillatory shear stress, we find the response of the system to be significantly shear banded, for a wide range of values of the amplitude $\gamma_0$ (or $\sigma_0$) and frequency $\omega$ of the imposed oscillation. Indeed, our results (in the glass phase $x<1$ at least) suggest that in the limit $\omega\to 0$, significant banding will be present for all imposed strain amplitudes in the nonlinear regime (with a smaller range of amplitudes implicated for larger frequencies).  We emphasize that this is true even though the model's underlying constitutive curve is purely monotonic, such that its steady state response to a steadily imposed shear of constant rate is incapable of supporting shear bands.  We attribute this to a repeated competition, within each cycle, of glassy aging and flow rejuvenation.

In the four strain-imposed protocols, the formation of shear bands in each half cycle appears closely associated with the presence of a stress overshoot in the elastic Lissajous Bowditch curve of stress as a function of strain, in close analogy to the transient shear banding associated with stress overshoot in shear startup studied previously~\cite{divoux2010,Divoux:2011b,Moorcroft2011,Moorcroft2013a}. Loosely and intuitively, therefore, we interpret LAOStrain (and the other strain-imposed protocols) in terms of a repeating series of forward and reverse shear startup flows. Likewise, in the stress-imposed protocol the formation of shear bands in each half cycle appears closely associated with a yielding process, just beyond the point at which the stress first exceeds the yield stress in the underlying constitutive curve. Again, this closely mirrors the transient shear banding associated with yielding following the imposition of a step stress studied previously~\cite{Divoux:2011a,Moorcroft2013a}. Loosely and intuitively, therefore, we interpret LAOStress in terms of a repeating sequence of positive and negative step stress experiments.

Our results suggest a possible generic predisposition of aging glassy materials to flow in a heterogeneous, shear banded manner when subject to large amplitude time-varying flows of even arbitrarily slow time-variation. It would be very interesting to investigate this suggestion further, both experimentally and in molecular simulations of glassy systems.
\begin{acknowledgments}
  The research leading to these results has received funding from the European Research Council under the European Union's Seventh Framework Programme (FP7/2007-2013) / ERC grant agreement number 279365. The authors thank Prof. P. Sollich for providing data to check our code in its homogeneous flow mode in Fig.~\ref{fig:comp_sollich}.
\end{acknowledgments}
%

\begin{thebibliography}{73}
\bibitem [{\citenamefont {{$1/\tau=\sum_{i=1}^{i=n} \sum_{j=1}^{j=m}\exp ( -
  (E_{ij} - k l_{ij}^2)/x)/(m n)$}}()}]{SM}%
  \BibitemOpen
  \bibfield  {author} {\bibinfo {author} {\bibnamefont
  {{$1/\tau=\sum_{i=1}^{i=n} \sum_{j=1}^{j=m}\exp ( - (E_{ij} - k
  l_{ij}^2)/x)/(m n)$}}},\ }\href@noop {} {\bibinfo  {journal} {where $E_{ij}$
  and $l_{ij}$ are energy trap depth and strain corresponding to an SGR element
  $ij$. $k=1$ in our units and is assummed to be same for all elements}\
  }\BibitemShut {NoStop}%
\end{thebibliography}

\end{document}